\documentclass{emulateapj}
\usepackage{apjfonts}
\slugcomment{\today}
\usepackage{natbib}
\usepackage{graphicx}
\usepackage{comment}
\usepackage{appendix}
\usepackage{subfigure}

\newif\ifcheck
\checktrue

\def\kms{\hbox{$~$km$~$s$^{-1}$}\,}
\def\arcdeg{\hbox{$^\circ$}\,}
\def\deg{\hbox{$^\circ$}\,}

\def\arcsec{\hbox{$^{\prime\prime}$}\,}
\def\ha{H$\alpha$\,}

\usepackage{color}

\shorttitle{Light Echo Spectroscopy of SN~1987A}
\shortauthors{Sinnott et al.}

\begin{document}

\title{Asymmetry in the outburst of SN~1987A detected using light echo
spectroscopy}

\author{B. Sinnott\altaffilmark{1}, 
	D. L. Welch\altaffilmark{1},
	A. Rest\altaffilmark{2},
	P. G. Sutherland\altaffilmark{1},
	M. Bergmann
}
\altaffiltext{1}{Dept. of Physics and Astronomy, McMaster University, Hamilton, Ontario, L8S 4M1, Canada}
\altaffiltext{2}{Space Telescope Science Institute, Baltimore, MD 21218, USA}

\begin{abstract}
We report direct evidence for asymmetry in the early
phases of SN~1987A via
optical spectroscopy of five fields of its light echo system. The light echoes
allow
the first few hundred days of the explosion to be reobserved, with different
position angles providing different viewing angles to the
supernova. Light echo spectroscopy therefore allows a direct
spectroscopic comparison of light originating from different regions of the
photosphere during the early phases of SN~1987A. Gemini
multi-object spectroscopy of the light echo fields shows fine-structure in the
\ha line as a smooth function of position angle on the near-circular light echo
rings. \ha profiles originating from the northern hemisphere of SN~1987A show
an excess in redshifted emission and a blue knee, while southern hemisphere
profiles show an excess of blueshifted \ha emission and a red knee. This
fine-structure is reminiscent of the ``Bochum event'' originally observed for
SN~1987A, but in an exaggerated form. Maximum
deviation from symmetry in the \ha line is observed at position angles
$16^{\circ}$ and
$186^{\circ}$, consistent with the major-axis of the expanding elongated ejecta.
The asymmetry signature observed in the \ha line smoothly diminishes as a function
of viewing angle away from the poles of the elongated ejecta. We propose an
asymmetric two-sided distribution of $^{56}$Ni most dominant in the
southern far quadrant of SN~1987A as the most probable explanation of the
observed light echo spectra. This is evidence that the asymmetry of
high-velocity $^{56}$Ni in the first few hundred days after explosion is
correlated to the
geometry of the ejecta some 25 years later.
\end{abstract}

\keywords{supernovae: individual(SN~1987A) --- supernova: general --- supernova
remnants --- line: profiles}

\section{Introduction}
\label{sec:intro}
A light echo (LE) occurs when outburst light from an event is scattered by
circumstellar or interstellar dust into the line of sight of the observer.
LE imaging has long been used as a powerful tool for studying the
three-dimensional dust structure surrounding supernovae (SNe)
\citep{crotts88,xu95,sugerman05,kim08}. In 1988, spectra of the inner
and outer LE rings of SN~1987A confirmed the resemblance to a maximum-light
spectrum \citep{gouiffes88,suntzeff88}. More recently, however, this technique
of targeted LE spectroscopy has been used on historical SNe to identify the
spectral type of the original outburst. After the serendipitous discovery of LEs
from ancient SNe in the Large Magellanic Cloud (LMC) during the SuperMACHO
project by \citet{rest05}, follow-up spectroscopy by \citet{rest08b} identified
the type of SN responsible
for the remnant SNR 0509-675. The spectral types of the Cas A and Tycho SNe have
also been identified using this method of targeted LE spectroscopy
\citep{rest08a,krause08casa,krause08tycho}. \citet{casaspec} used LE
spectroscopy to detect asymmetries in the outburst of
the Cas A SN, while \citet{eta} recently obtained LE spectroscopy from the
``Great Eruption'' of $\eta$ Carinae. We refer the reader to \citet{lereview}
for a review of LE spectroscopy emphasizing these more recent results. Given the
acceleration of the field of LE spectroscopy in recent years, a detailed study
of spectra of the well-known LE system of SN~1987A can provide a foundation for
future studies, as well as provide new insight into the explosion of SN~1987A.

In the case of historical LEs, one has to use the photometric and spectroscopic
history of a different SN to model the LE spectrum \citep[see, e.g.][where the
Cas A outburst is modeled with the lightcurve and spectra of
SN~1993J and SN~2003bg]{casaspec}. For SN~1987A, however, the exact spectral and
photometric history of the LE source is known with high-precision, allowing
observed LE spectra to be compared unambiguously to an isotropic scenario. In
addition to acting as a test bed for LE spectroscopy theories, the near-circular
LE rings of SN~1987A allow the original outburst to be probed for asymmetries in
an entirely direct way. Different position angles (PAs) on the LE system probe
different viewing angles from which we can view the time-integrated spectrum of
the original event.

Observations as well as theoretical simulations indicate that core-collapse
SNe are asymmetric in nature. Polarization measurements show
deviations from spherical symmetry in all core-collapse SN with sufficient data
\citep{wang08}, while recent simulations in two and three spatial dimensions
show large deviations from symmetry
\citep[e.g.,][]{hammer10,gawryszczak10,muller12}. Coupled
with the fact that state of the art spherically symmetric core-collapse
simulations in one spatial dimension fail at producing an explosion for
$>10$M$_{\sun}$ projenitors \citep{burrows12}, multi-dimensional
physics 
such as the standing accretion shock instability
\citep[SASI;][]{blondin03} may play a key role in understanding the explosion
mechanism. As a new method for observing asymmetries, LE spectroscopy of SNe
may be able to provide new insight into the origin of SN
asymmetries and their relation to the core-collapse explosion mechanism.

SN~1987A was a peculiar Type II SN with a blue supergiant progenitor located 
in the LMC \citep[see, e.g.,][and references therein for a review of SN~1987A and its progenitor]{arnett89}. SN~1987A was known to be an asymmetric SN. Early polarization
measurements \citep[e.g.,][]{jeffery87,bailey88,cropper88} and an elongated
initial speckle image \citep{papaliolios89} suggested a
non-spherical event. Fine structure in the \ha line (the
``Bochum event'') at
20-100 days after the explosion as well as
redshifted emission lines at more
than 150 days after explosion provide strong evidence for radial-mixing of heavy
elements into the upper envelope
\citep{hanuschik87,phillips89,spyromilio90}. Updated models of the
bolometric lightcurve at early epochs also require radial-mixing of $^{56}$Ni
\citep{shigeyama90,utrobin04}. Direct HST imaging by
\cite{wang02} showed an elongated remnant ejecta, claimed to be bipolar and
showing evidence for a jet-induced explosion. Recent integral field
spectroscopy of the ejecta by \citet{kjaer10} showed a prolate structure for the
ejecta oriented in the plane of the equatorial ring, arguing against a
jet-induced explosion.

Here we present detailed imaging and spectroscopy of the SN 1987A LE system. We
infer the relative contributions of the different epochs of the SN to the
observed LE spectrum by modeling images of the LE, as demonstrated by
\citet{leprofile}. We fit the observed LE spectra and use specific examples to
illustrate the models ability to correctly interpret the observations. We then
use the LE spectra to probe for asymmetries in the explosion of SN~1987A and
compare to previous observations of asymmetry.
\section{Observations and Reductions}
\label{sec:observations}
\subsection{Imaging}
\label{sec:observations:imaging}
Imaging of the SN~1987A LE system was performed under the SuperMACHO Project
microlensing survey \citep{supermacho} using the CTIO 4m Blanco telescope. The
survey monitored the central portion of the LMC for five seasons
beginning in 2001 using the 8K $\times$ 8K MOSAIC imager (plus atmospheric
dispersion corrector) with the custom ``VR'' filter ($\lambda_{c}=625$ nm,
$\delta \lambda=220$ nm). Exposure times were between 150 and 200 s. We have
continued to monitor the field containing SN~1987A and its LEs (field sm77)
since the survey ended. Data reduction and difference images were performed
using the ESSENCE/SuperMACHO pipeline {\it
photpipe} \citep{supermacho,garg07,miknaitis07}. A stacked and
mosaiced difference image of the LE system is shown in
Figure~\ref{fig:slit_locations}. The near-circular rings illuminate three
general dust structures: a smaller structure $\sim85$~pc in front of SN~1987A,
only visible in the south; a near-complete ring at $\sim185$~pc and brightest
in the north-east; as well as a larger and fainter near-circular illumination
at $\sim400$~pc in front of the SNR. All three of these structures have been
mapped previously in detail by \citet{xu95} and can be referred to as three
``sheets'' of ISM dust roughly in the plane of the sky. However, it is
important to note that the LE flux observed in
Figure~\ref{fig:slit_locations} is due to dense filamentary
structure within the general ``sheets'' of ISM dust. The
physical properties of the scattering dust filaments (inclination and thickness)
can vary greatly within what appears to be a uniform ``sheet'' of dust. This
distinction is important to make when modeling the LE photometry and
spectroscopy.
\subsection{Gemini Spectroscopy}
\label{sec:observations:gemini}
\begin{figure}[]
\epsscale{1.0}
\plotone{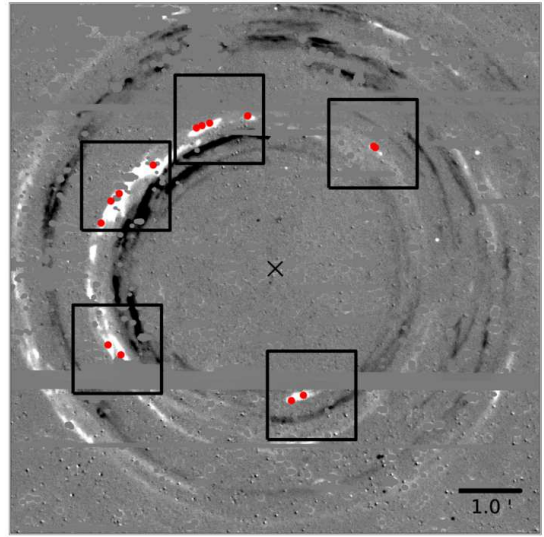}
\caption[]{Mosaiced SuperMACHO difference image of August 2006 and January 2002
epochs of the SN~1987A LE system. The locations of the 14 GMOS slits
appear
in red, while the black cross indicates the location of SNR. Slitlets are
not to scale. The black squares collect LEs which were observed in the same MOS
field. North is up and east is to the left.
\label{fig:slit_locations}}
\end{figure}

We obtained multi-object optical spectroscopy (MOS) for five fields of the
SN~1987A echo system in the 2006B term, using the R400 grating and GG455
blocking filter on the Gemini Multi-Object Spectrograph (GMOS) on Gemini-South.
SuperMACHO or GMOS preimages were differenced with previous
SuperMACHO images to establish the precise location of the echo system, allowing
the design of GMOS masks. The locations of five MOS fields and the 14 1.0\arcsec
wide LE slitlets are shown in Figure~\ref{fig:slit_locations}.

The nod and shuffle mode of GMOS was used to best isolate the diffuse LE signal
from the nebular sky background. To ensure clear sky in the offset
position, the telescope was nodded off-source several degrees every 225
s. On-source integration times for the brightest echo fields were 30 minutes,
and 90 minutes for the fainter, more diffuse echoes. During each night of
observations, flat field
and CuAr spectral calibration images were taken before or after the science
images. We also acquired dark images with the same nod and shuffle parameters
used during integration. These specialized dark images allow features
associated with
charge-traps in the GMOS CCDs to be masked out during reduction.

The observations result in LE spectra from SN~1987A with a
spectral range of 4500-8500 \AA, contaminated by strong emission lines from the
nebular region, as well as any stellar continuum entering the slitlets.
The P Cygni \ha line seen in the spectra is further degraded due to six
strong nebular emission lines ([NII] $\lambda$6548, \ha $\lambda$6563,
[NII] $\lambda$6585, HeI $\lambda$6678, [SiII] $\lambda$6717,
[SiII] $\lambda$6731). Signal from the emission peak of the P Cygni \ha line
is therefore lost to nebular emission, and this region of the profile cannot be
interpreted with confidence. In particular, the location and flux of the
emission peak cannot be determined without interpolations which we found
introduced large uncertainties.

The nebular contamination can be greatly reduced since the LE
system is expanding superluminally on the sky (as seen in
Figure~\ref{fig:slit_locations}). We
therefore have the unique opportunity in astronomy to observe only the
background sky signal at the \emph{same location on the sky} as the initial
sky+object
observation. Sky-only observations were taken in the 2009A and 2009B semesters
for the five MOS fields. The spectroscopy was carried out using the same GMOS
configuration as the previous sky+object 2006B observations, with 30 minutes
on-source integration times in nod and shuffle mode. The spectrophotometric
standard star LTT 3864 was observed during the 2009A observations.
\subsection{Spectra Reduction}
\label{sec:observations:reductions}
All spectra were reduced using IRAF\footnotemark~and the Gemini IRAF package,
\footnotetext{IRAF is distributed by the National Optical Astronomy
Observatory,
which is operated by the Association of Universities for Research in Astronomy,
Inc., under cooperative agreement with the National Science Foundation.}
with bias subtraction and trimming done on each CCD individually. Science images
were dark subtracted using the special nod and shuffle darks, with each CCD
being cleaned of cosmic rays
using the Laplacian rejection algorithm LACOSMIC \citep{vandokkum01}. Nod and
shuffle sky subtraction was performed on each science exposure using the
\emph{gnsskysub} task. Individual slitlets were automatically cut from the
wavelength calibrations frames, however this procedure required manual tweaking
to achieve the best cutting possible. A wavelength solution was found for each
slitlet in the
calibration frame. Science frames were flattened, CCDs mosaiced together, and
slitlets cut from the images. The wavelength solution for each slitlet was then
applied to the corresponding science spectrum in two dimensions. Since the LE
signal is so weak, determining an extraction aperture is very difficult.
Instead, we collapsed the entirety of each slitlet to one dimension using a
block average.

No suitable spectrophotometric standard star was observed during the 2006B
observing semester, nor did a suitable standard star matching our
unique instrument configuration exist in the data archive nearby temporally.
However,
the spectrophotometric standard LTT 3864 was observed during the 2009A
observations. We therefore used the 2009A standard to perform the flux
calibration for all of the 2006 and 2009 spectra. While this leads to large
errors in absolute flux calibration, our analysis is independent of absolute
flux levels. Instead, the purpose of the flux calibration
used here is to remove the effect of CCD sensitivity from the spectra, which is
relatively stable in time. A LE spectrum is a weighted average of many epochs
from the outburst. Determining the continuum of such a spectrum is therefore
difficult, and the alternative analysis procedure of removing the
continuum from all of our spectra would have had a larger error than that of
the flux calibration.

The sky-only spectra from 2009 were subtracted in one dimension from the
2006 LE spectra for each MOS slit. This procedure was
performed interactively, adjusting the scale and wavelength of the sky spectrum
in small increments using the IRAF
task \emph{skytweak} to minimize sky
subtraction residuals in the final spectra. Since we focus our science analysis
on the \ha line, we performed the sky subtraction such that the subtraction
residuals in the P Cygni profile of \ha were minimized. All spectra were
Doppler-corrected, adopting a LMC radial velocity of 286.5\kms
\citep{meaburn95}.
\section{Analysis}
\label{sec:analysis}
\subsection{Light Echo Spectra}
\label{sec:spectra}
\begin{figure}[h]
\epsscale{1.3}
\plotone{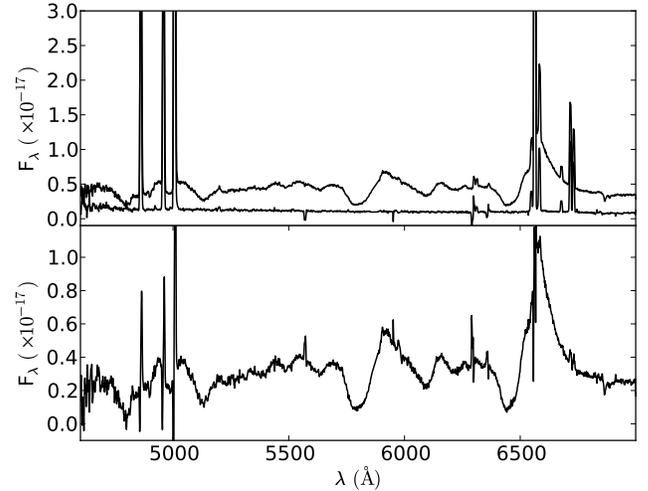}
\caption[]{Upper Panel: Example LE+sky spectrum from 2006 shown with the
corresponding sky-only nebular spectrum from 2009. Lower Panel: The result of
difference spectroscopy, showing the LE-only spectrum along with remaining
nebular residuals from the subtraction. The strength of the \ha emission peak
can now be estimated in the difference spectrum.
\label{fig:lespec}}
\end{figure}

The upper panel of Figure~\ref{fig:lespec} shows a reduced LE spectrum
from the 2006 observations along with the underlying sky-only nebular spectrum
from 2009. The high-velocity P Cygni profiles including \ha resemble a Type II
SN spectrum at maximum light, confirming the LE spectra probe the original
outburst light of SN~1987A. Strong, narrow nebular
lines not associated with the SN outburst dominate the spectrum around
5000\AA~and the \ha line at 6563\AA.
The result of performing difference spectroscopy is shown in the lower panel of
Figure~\ref{fig:lespec}, where the sky-only signal has been subtracted from the
2006 echo spectrum. Flux from the emission component of the \ha line is
recovered that was initially lost to nebular contamination.  In
the majority of our
spectra we are able to resolve the \ha emission peak without loss due to the
nebular lines.
\subsection{Modeled Isotropic Spectrum}
\label{sec:model}
\begin{figure}[h]
\epsscale{1.2}
\plotone{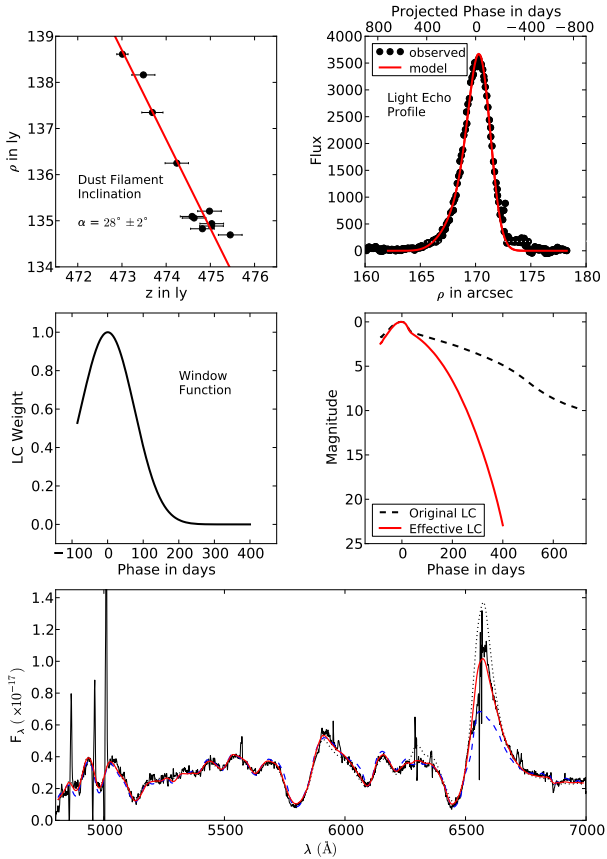}
\caption[]{Summary of modeling procedure for each LE spectrum. The
inclination of the dust filament is measured (upper left panel), allowing the
shape of the LE profile to be modeled to determine the dust width (upper right
panel). The properties of the slit determine the window function containing the
relative contribution from each epoch (middle left panel), which when multiplied
by the original lightcurve gives the effective lightcurve for this observation
(middle right panel). We integrate the historical spectra of SN~1987A,
flux-weighted with the effective lightcurve, and take into account
wavelength-dependent scattering and reddening to obtain a fit between the LE
model and the observed spectrum (lower panel). The lower
panel plot shows the model LE fit (solid red) along with the attempted fit with
the maximum light spectrum of SN~1987A (dashed blue) and the full
lightcurve-integrated spectrum (dotted black).
\label{fig:modelsummary}}
\end{figure}
To search for asymmetries in SN~1987A using spectroscopy of its LEs,
we need to first consider what defines asymmetry in observed spectra. As
shown in \citet{leprofile}, multiple LE spectra from the same transient
source, with differing line strength ratios, does not necessarily imply an
asymmetry in the outburst light. An observed LE spectrum
will depend
strongly on the physical properties of the scattering dust as well as the
instrument configuration and seeing conditions at time of observation.
Specifically, the dominant epochs of the original outburst probed by a LE
spectrum can change with the above properties. Relative
comparisons of LE spectra can therefore easily lead to false-detections
of asymmetry if the above factors are not taken into consideration. Each
LE spectrum should instead be compared to a single isotropic spectrum modeled
for each observation. We define
our isotropic model
spectrum as the original outburst of SN~1987A, as it would have been
spectroscopically observed if scattered by dust corresponding to the observed
LE. This provides a direct probe of asymmetry since differences
between observed and model spectra represent deviations from the historically
observed outburst of SN~1987A.

A LE observed on the sky has a flux profile (flux versus $\rho$, where
$\rho$ is the distance on the sky from the SNR, see upper right
panel of Figure~\ref{fig:modelsummary}) that is the
projected lightcurve of the source event stretched or compressed depending
on the inclination of the scattering dust. The width of the scattering dust and
the point spread function (PSF) further distort the observed LE. The
dust inclination is measured through imaging the LE at multiple epochs and can
be measured prior to spectroscopy. The PSF is also known, allowing the width of
the scattering dust filament to be determined through fitting the LE flux
profile.

The inclination, $\alpha$, is obtained by monitoring the apparent motion of the
LE on the sky. Positive values of $\alpha$ correspond to scattering dust sheets
tilted out of the plane of the sky, from the positive $\rho$ axis toward the
negative $z$ axis, where $z$ is the distance in front of the remnant. Using the
SuperMACHO database of difference images around SN~1987A, we are able to monitor
the apparent motion of
the echo system over seven years from 2002 to 2008. For each difference image
and each slit location in
Figure~\ref{fig:slit_locations}, we measure the one dimensional profile of
the difference flux as a function of distance away from the SNR, $\rho$.
Carefully monitoring the superluminal apparent motion in this way allows us to
determine the inclination of the scattering dust for each slit location, as
described in \citet{leprofile} (see, e.g., upper left panel of
Figure~\ref{fig:modelsummary}). We find the
apparent motion (and therefore
inclination) is typically stable over a one year period. We find typical
uncertainties in the inclination are less than $5^{\circ}$ for our 14 
dust locations. \citet{leprofile} showed that, for the case of SN~1987A, changes
in
inclination on the order of $10^{\circ}$ do not alter the final modeled
spectrum in a significant way. With the inclination known, the photometric
history of SN~1987A
\citep{87aphoto_hamuy88,87aphoto_suntzeff88,87aphoto_hamuy90} can be used to
model the one dimensional flux profile and determine the best-fitting scattering
dust width. The upper right panel of Figure~\ref{fig:modelsummary} shows an
example fit to the LE profile.

The above procedure determines the properties of the scattering dust for each
LE location. We also take into account the location, orientation, and size
of the spectroscopic slit to compute a window function that is unique to each
LE observation (see, e.g., middle left panel of Figure~\ref{fig:modelsummary}).
This window function is the relative contribution from each epoch of the
lightcurve when observed through the slit. The slit offset, $\Delta\rho$,
measures the offset between the peak of the LE profile and the location of the
slit on the sky. This is an important parameter, as it shifts the window
function in time, systematically probing later or early epochs in the spectrum
if the slit is positioned closer or further away from the SNR, respectively.  

Multiplying the window
function with the original lightcurve of SN~1987A, we determine an
\emph{effective lightcurve} corresponding to each observed LE spectrum (see,
e.g., middle right panel of Figure~\ref{fig:modelsummary}). The isotropic model
spectrum is then the integration of the historical spectra of SN~1987A weighted
with the effective lightcurve. We refer the reader to \citet{leprofile} where
our model is described in
detail with SN~1987A examples. To integrate the historical spectra of SN~1987A,
we use the integration method described in \citet{rest08b}, with the spectral
database from SAAO and CTIO observations
\citep{87aspectra1,87aspectra2,87aspectra3,87aspectra4,87aspectra5,87aspectra6,
87aspectra7,phillips88,phillips90}.

The wavelength-dependent effect of scattering by dust grains is taken into
account when fitting the isotropic model to our observed LE spectra. We
calculate the integrated scattering function for each LE (i.e. scattering
angle) using the method described in \citet{lereview}, using the \emph{``LMC
avg''} carbonaceous-silicate grain model of \citet{weingartner01}. We
then fit the model spectrum to the
observed LE spectrum (see, e.g., lower panel of Figure~\ref{fig:modelsummary})
varying three parameters: the reddening (E(B-V) assuming R$_v=3.2$), a
normalization constant, and a small flux offset (to account for small errors in
sky subtraction).

Figure~\ref{fig:modelsummary} summarizes the
modeling procedure outlined above and in \citet{leprofile} that occurs for each
LE spectrum. The lower panel shows the final isotropic model spectrum in red,
which fits the features in the observed LE spectrum very well. To highlight the
importance of the modeling procedure, we also attempt to fit the LE spectrum
with a spectrum of SN~1987A near maximum light (dashed blue line), and an
integrated
spectrum obtained by flux-weighting with the original lightcurve of SN~1987A
(dotted black line). The maximum light spectrum cannot reproduce the strength
of the
observed \ha emission. Conversely, the full lightcurve-integrated spectrum has
an excess in \ha emission and a lower \ha velocity in the absorption minimum.
Only by integrating using the modeled window function (solid red line) are we
able
to obtain a good fit to the \ha profile. Below we highlight the importance of
the LE modeling by considering two scenarios.
\subsubsection{Dust-Dominated Scenario}
\label{sec:mode_f2_example}
\begin{figure}[h]
\epsscale{1.20}
\plotone{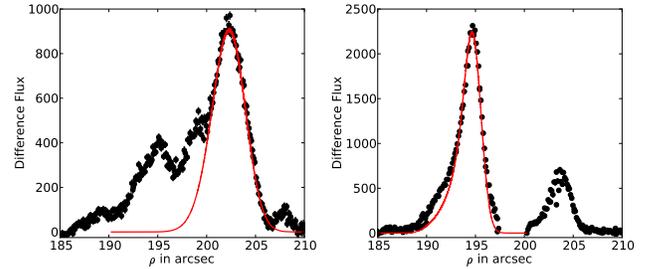}
\caption[]{Observed LE profiles and corresponding best-fit models (red line)
for LE113 (left) and LE117 (right). The larger width and symmetric shape of the
LE113 profile on the sky is due to the dust filament being inclined
$\sim50^{\circ}$ closer to the tangential to the line-of-sight than the LE117
dust filament, compressing the entire lightcurve on the sky and resulting in a
more symmetric point-like LE profile.
\label{fig:dust_example_f2}}
\end{figure}

Both the physical properties of the scattering dust as
well as the configuration of the observation affect the observed LE spectrum.
Here
we highlight the effect of the scattering dust. Table~\ref{tab:modelresults}
shows the
observed inclinations and best-fitting dust widths for the 14 LE
locations shown
in Figure~\ref{fig:slit_locations}. Figure~\ref{fig:dust_example_f2} shows two
examples of observed flux profiles, corresponding to the two LEs in
Figure~\ref{fig:slit_locations} at PA~$\sim115^{\circ}$, LE113 and LE117. Note
the numbers in the names of the LEs correspond to the PA of the LE with respect to the
SNR. Both LEs are at essentially the same PA along the echo
system, so probe the same viewing angle onto the SN. However, as shown in
Figure~\ref{fig:slit_locations} the two slits are placed on physically
distinct dust filaments. The bright filament in LE113 is 53~ly closer to the
observer along the line of sight than the bright LE117 filament, while both
are 3~ly thick. The observed widths and shapes of the LE profiles in
Figure~\ref{fig:dust_example_f2} are also different, with LE113
having a larger width on the sky and a more symmetric shape. This is explained
entirely by the $\sim50^{\circ}$ difference in the inclinations of the two
dust filaments. The highly inclined dust
filament for LE113 is within $\sim15^{\circ}$ of the tangential to the
scattering ellipsoid and compresses the lightcurve on the sky, causing the LE113
profile to resemble a point-like source with a broad, symmetric profile. The
lower
inclination of the LE117 dust filament preserves the shape of the lightcurve
with a broad increase in flux at small $\rho$ values (the long decay of the
SN lightcurve) and a sharp decrease in flux at large $\rho$ values (the short
rise of the SN lightcurve). It should be noted that the LE113 profile in the
left panel of Figure~\ref{fig:dust_example_f2} shows dust substructure to the
left of the main LE peak. The
additional contribution from the secondary peak is weak pre-maximum flux and
we found the effect on the \ha emission strength in the integrated spectra to
be $<4\%$. However, we do take substructure in the dust into
account in the extreme case of LE186 as shown
in Section~\ref{sec:asymmetry:ha} below. The effect of dust substructure is
considered more formally in the Appendix.

\begin{figure}[h]
\epsscale{1.2}
\plotone{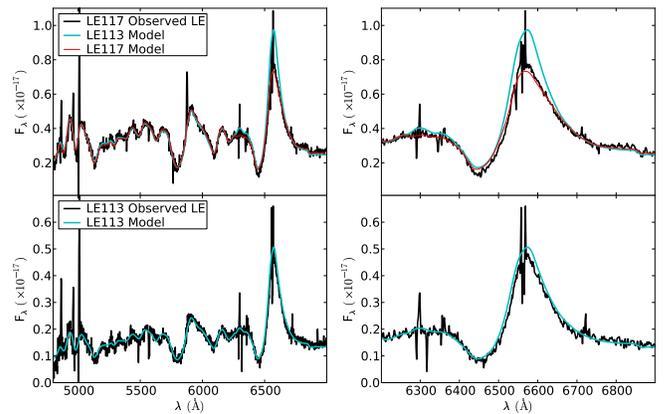}
\caption[]{The upper left panel shows the observed LE spectrum of LE117 with its
corresponding isotropic model, along with the incorrect model of the nearby dust
filament LE113. Only the matching LE117 model can fit the observed LE spectrum,
while the lower left panel shows the LE113 observed LE spectrum can be fit with
its corresponding model. The right panels show enlargements of the \ha profile
in both cases.
\label{fig:spec_example_f2}}
\end{figure}

The modeled isotropic spectra are shown in Figure~\ref{fig:spec_example_f2},
with the upper left panel showing the observed LE spectrum for the LE117 slit
(black), the corresponding LE117 model (thin red), and the non-matching
LE113 model (thick cyan) derived using the larger inclination. Since the
dust inclination of
LE113 compresses the lightcurve on the
sky, a much larger range of epochs span the width of the slit resulting in a
wide window function compared to the LE117 slit (Table~\ref{tab:modelresults}
lists the approximate range of epochs probed by each LE). The LE113 model
therefore has
an excess of \ha emission from the inclusion of late-time nebular epochs. The
LE117 model is able to correctly match the observed LE117 \ha strength. The
lower panel of Figure~\ref{fig:spec_example_f2} shows the same LE113 model with
the corresponding LE113 observed LE spectrum, showing good agreement. The fact
that two distinct LEs with differing line strengths and differing dust
properties show the same result -- that the observed LE spectra
from this viewing angle can be modeled with an isotropic historical spectrum
-- shows the interpretation of LE spectroscopy described here and in detail in
\citet{leprofile} is correct.
\begin{deluxetable*}{lrrrrrrr}
\tabletypesize{\scriptsize}
\tablecaption{
\label{tab:modelresults}}
\tablehead{
\colhead{LE ID\tablenotemark{a}} &
\colhead{PA\tablenotemark{b}} &
\colhead{$z_{0}$\tablenotemark{c}} &
\colhead{$\theta$\tablenotemark{d}} &
\colhead{$\alpha$\tablenotemark{e}} &
\colhead{$\sigma_{d}$\tablenotemark{f}} &
\colhead{$\Delta \rho_{\rm offset}$\tablenotemark{g}} &
\colhead{Epochs\tablenotemark{h}} \\
\colhead{} &
\colhead{(\arcdeg)} &
\colhead{(ly)} &
\colhead{(\arcdeg)} &
\colhead{(\arcdeg)} &
\colhead{(ly)} &
\colhead{(\arcsec)} &
\colhead{(days)}
}
\startdata
LE016 &   15.8 & 428.91 & 17.1 &  22 $\pm$   6&  4.49 $\pm$ 0.03  & -0.35   &
0-180\\
LE029 &   29.2 & 473.13 & 16.3 &  28 $\pm$   2&  4.49 $\pm$ 0.03 & -0.19 &
5-200\\
LE032 &   31.8 & 483.02 & 16.2 & -46 $\pm$   2&  2.71 $\pm$ 0.05 & -0.27 
&30-180\\
LE034 &   33.9 & 486.53 & 16.1 & -16 $\pm$   2&  3.37 $\pm$  0.15 &  +0.26 
&0-140\\
LE053 &   53.0 & 505.05 & 15.8 &  38 $\pm$   7&  5.26 $\pm$ 0.03  & -0.18  
&0-215\\
LE066 &   66.3 & 590.39 & 14.7 &  11 $\pm$   2&  4.50 $\pm$ 0.55  &  +1.05 
&0-80\\
LE069 &   69.3 & 624.64 & 14.3 &  66.8 $\pm$ 0.2&  1.80 $\pm$ 0.07   &  +0.57
&0-110\\
LE076 &   76.4 & 648.50 & 14.0 &  11 $\pm$   4&  0.60 $\pm$ 0.22  & -0.08  
&40-135\\
LE113 &  112.5 & 671.04 & 13.8 &  79 $\pm$   1&  3.06 $\pm$ 0.09  & -0.39  
&0-330\\
LE117 &  116.8 & 618.40 & 14.3 &  33 $\pm$   5&  3.10 $\pm$ 0.39  & -0.14  
&35-155\\
LE180 &  180.2 & 289.69 & 20.7 & -49 $\pm$   2&  4.53 $\pm$ 0.06  & +0.73  
&0-290\\
LE186 &  185.6 & 270.34 & 21.3 &  70 $\pm$   3&  3.52 $\pm$ 0.18  & -1.00  
&0-420\\
LE325 &  325.3 & 369.67 & 18.4 &  -6 $\pm$  11&  1.13 $\pm$ 0.16  &  +0.18 
&20-120\\
LE326 &  326.1 & 370.33 & 18.4 &  15 $\pm$   3&  2.24 $\pm$ 0.51  &  +0.21 
&0-130\\
\enddata
\tablenotetext{a}{Naming convention corresponds to position angle of LE slit
with respect to SNR.}
\tablenotetext{b}{Position angle of the scattering dust with respect to the
SNR.}
\tablenotetext{c}{Distance along line of sight from SNR to scattering dust.}
\tablenotetext{d}{Scattering angle.}
\tablenotetext{e}{Inclination of scattering dust with respect to the plane of
the sky. $\alpha$ increases from the positive $\rho$ axis towards the negative
$z$ axis.}
\tablenotetext{f}{Average best-fitting width of scattering dust sheet over the
length of the spectroscopic slit.}
\tablenotetext{g}{Offset between center of slit and peak LE flux, where
positive values indicate slit further from SNR than peak echo flux.}
\tablenotetext{h}{Approximate range of epochs probed by the LE. Corresponds to
epochs in window function in which the relative contribution contribution is
$>50\%$ (prior to flux-weighting) in the LE
integration.}
\end{deluxetable*}
\subsubsection{Observation-Dominated Scenario}
\label{sec:mode_f5_example}
\begin{figure}[h]
\epsscale{1.2}
\plotone{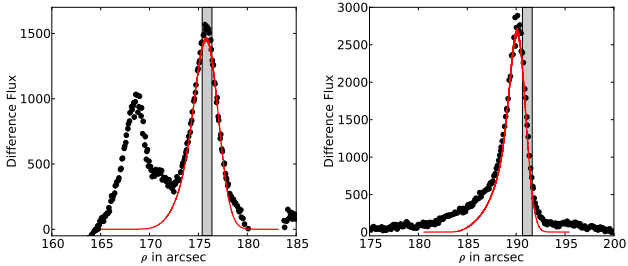}
\caption[]{Observed LE profiles and corresponding best-fit models (red line)
for LE053 (left) and LE066 (right). The grey shaded regions indicate the
location of the spectroscopic slit. The slit offset from the LE peak in the
LE066 LE produces an observed spectrum that is heavily weighted towards the
earlier epochs of the outburst.
\label{fig:dust_example_f5}}
\end{figure}

\begin{figure}[h]
\epsscale{1.2}
\plotone{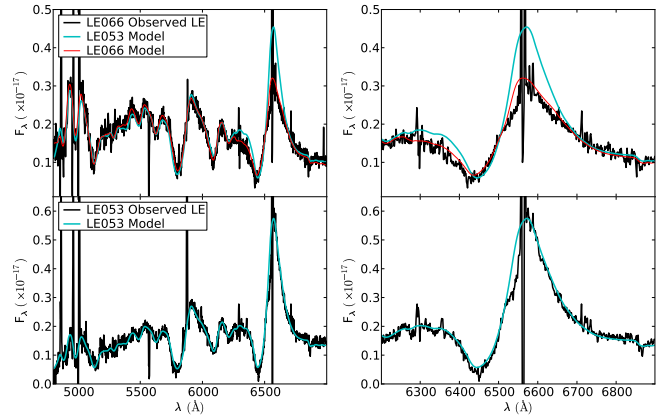}
\caption[]{The upper left panel shows the observed LE spectrum of LE066 with its
corresponding isotropic model, along with the incorrect model of the nearby dust
filament LE053. Only the higher velocity LE066 early-epoch model is a good fit
for the observed LE spectrum. The lower left panel shows the LE053 observed LE
spectrum which probes later epochs of the explosion. The lower velocity and
larger \ha emission-to-absorption ratio of the LE053 model is a good fit to the
observed spectrum. The right panels show enlargements of the \ha profile
in both cases.
\label{fig:f5s21_comparison}}
\end{figure}

Here we consider a scenario where the properties of the observation
(specifically the slit location) are the dominant factors in two very
different observed LE spectra that have similar viewing angles onto the
photosphere. Figure~\ref{fig:dust_example_f5} shows the LE profiles of LE053 and
LE066, corresponding to the slits at PA$\sim60^{\circ}$ in
Figure~\ref{fig:slit_locations}. The gray shaded region indicates the location
of the slit for both profiles. The slit of LE066 was placed $\sim1\arcsec$
farther away from the SNR than the peak of the LE. This was unintentional since
the location of the LE peak may not be very well known when
making the MOS masks. The offset results in very different observed LE
spectra for the
two profiles, as shown in Figure~\ref{fig:f5s21_comparison}. When compared to
the observed LE spectra of Figure~\ref{fig:spec_example_f2} and LE053 in
Figure~\ref{fig:f5s21_comparison}, the spectrum of LE066 looks more like an
early-time spectrum with a lower \ha emission-to-absorption ratio and a higher
temperature continuum. Since the slit is centered on the pre-maximum light
portion of the LE profile, LE066 does not probe epochs $>100$~days after maximum
light, while LE053 probes epochs out to $\sim300$~days after maximum light. As
in
the previous example, both model isotropic spectra are good fits to their
observed LE spectra. However, the model of LE053 has a lower \ha absorption
velocity and a larger \ha emission-to-absorption ratio, making it a poor fit to
the LE066 observed LE spectrum.

The above dust- and observation-dominated examples highlight the danger in
comparing LE spectra
without sufficient modeling. Without the corresponding models for the LE053
and LE066 observed spectra, one might make a false-detection of asymmetry: the
LE066 spectrum has a larger \ha velocity and excess in emission compared to the
LE053 spectrum. If the two LEs occurred at opposite PAs this asymmetry claim
would be even more tempting to make.
\newpage
\subsection{Evidence for Asymmetry in SN~1987A}
\label{sec:asymmetry}
Each PA on the LE system of SN~1987A represents a distinct viewing angle with
which to view the original outburst. LEs originating to the north
of SN~1987A probe outburst light originating mainly from the northern portion of the
photosphere. Therefore, by observing the LEs as a function of
increasing PA, we are able to view the outburst of SN~1987A as a
function of north to south lines of sight. The scattering angles probed by the LE 
ring are listed in Table~\ref{tab:modelresults} and lead to typical opening 
angles of $\sim35^{\circ}-40^{\circ}$.

\begin{figure}
\epsscale{1.25}
\plotone{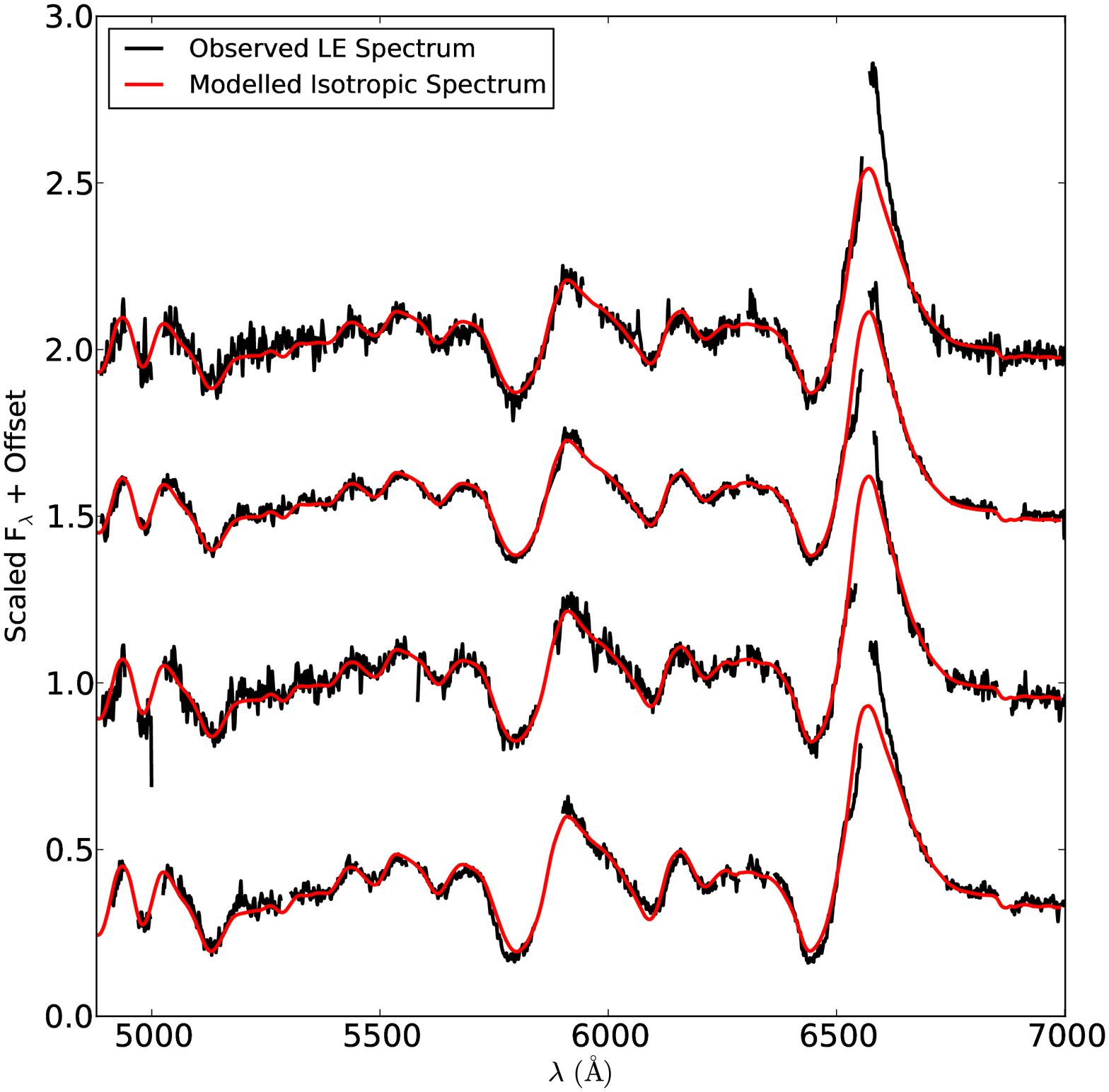}
\caption[]{Observed LE spectra with corresponding modeled isotropic spectra
for field at PA $\sim30\arcdeg$ in Figure~\ref{fig:slit_locations}.
Spectra are plotted from top to bottom as a function of increasing position
angle as shown in Figure~\ref{fig:slit_locations}: LE016, LE029, LE032, LE034.
Gaps in the observed spectra correspond to areas contaminated by
sky-subtraction residuals.
\label{fig:6}}
\end{figure}

\begin{figure}
\epsscale{1.25}
\plotone{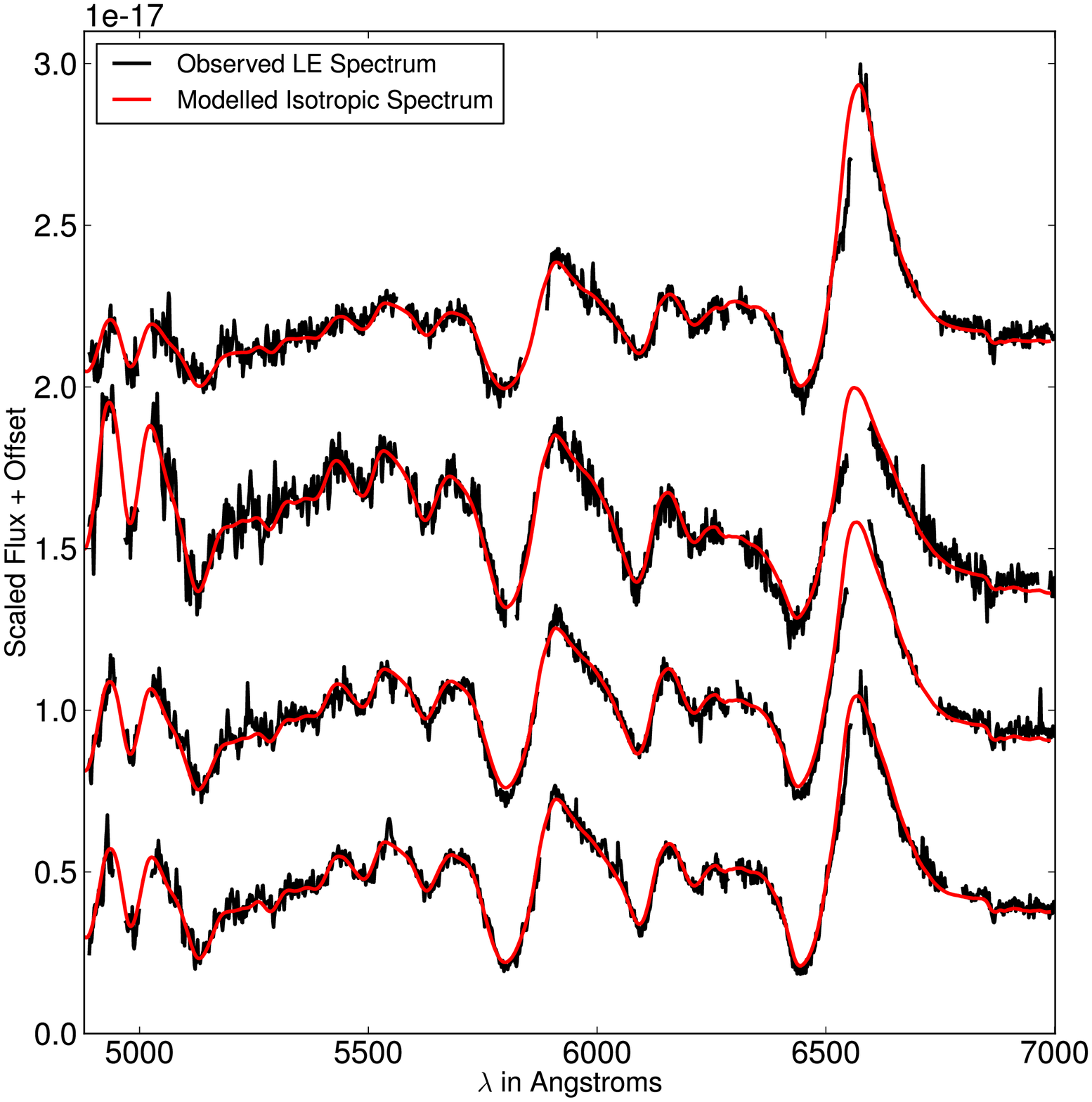}
\caption[]{Observed LE spectra with corresponding modeled isotropic spectra
for field at PA $\sim65\arcdeg$ in Figure~\ref{fig:slit_locations}.
Spectra are plotted from top to bottom as a function of increasing position
angle as shown in Figure~\ref{fig:slit_locations}: LE053, LE066, LE069, LE076.
Gaps in the observed spectra correspond to areas contaminated by
sky-subtraction residuals.
\label{fig:5}}
\end{figure}

\begin{figure}
\epsscale{1.25}
\plotone{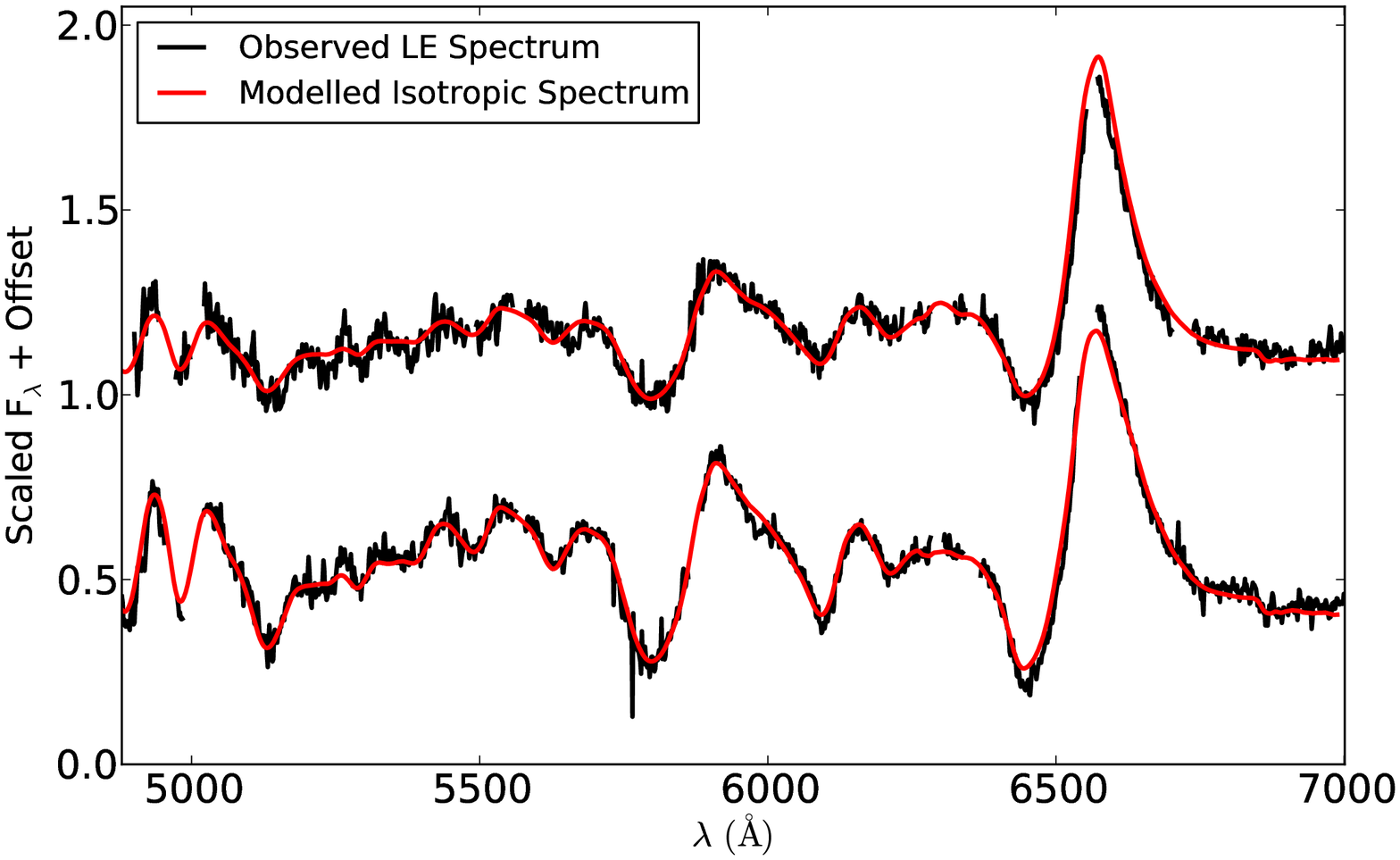}
\caption[]{Observed LE spectra with corresponding modeled isotropic spectra
for field at PA $\sim115\arcdeg$ in Figure~\ref{fig:slit_locations}.
Spectra are plotted from top to bottom as a function of increasing position
angle as shown in Figure~\ref{fig:slit_locations}: LE113, LE117. Gaps in the
observed spectra correspond to areas contaminated by
sky-subtraction residuals.
\label{fig:2}}
\end{figure}

\begin{figure}
\epsscale{1.25}
\plotone{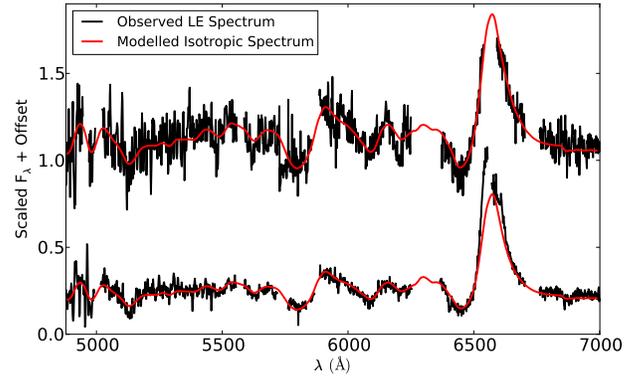}
\caption[]{Observed LE spectra with corresponding modeled isotropic spectra
for field at PA $\sim185\arcdeg$ in Figure~\ref{fig:slit_locations}.
Spectra are plotted from top to bottom as a function of increasing position
angle as shown in Figure~\ref{fig:slit_locations}: LE180, LE186. Gaps in the
observed spectra correspond to areas contaminated by
sky-subtraction residuals. LE180 (upper spectrum) represents an example where
our LE fitting algorithms fail to extract meaningful results due to the
complexity and low signal-to-noise of both the spectrum and the LE flux profile.
\label{fig:3}}
\end{figure}

\begin{figure}
\epsscale{1.25}
\plotone{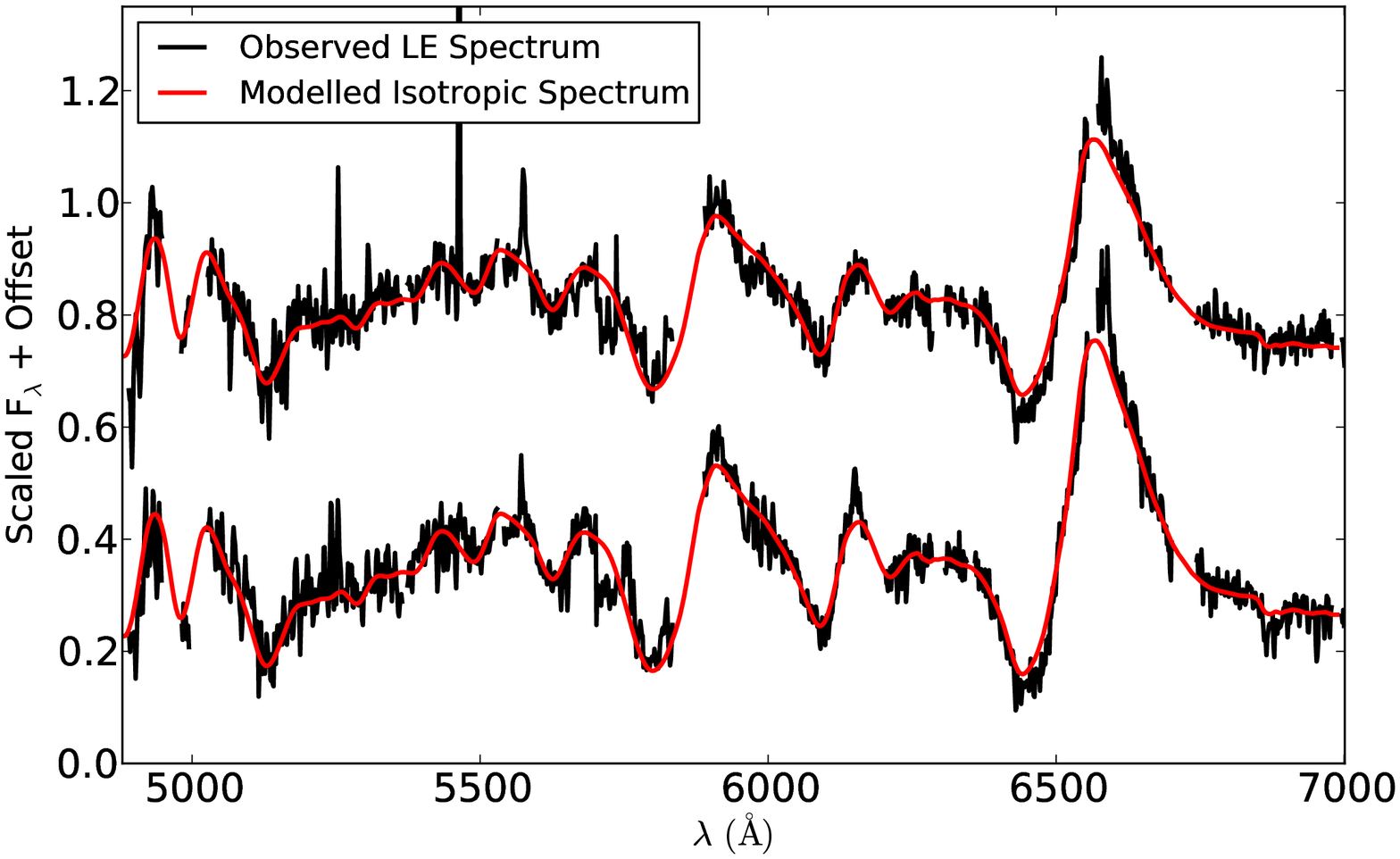}
\caption[]{Observed LE spectra with corresponding modeled isotropic spectra
for field at PA $\sim325\arcdeg$ in Figure~\ref{fig:slit_locations}.
Spectra are plotted from top to bottom as a function of increasing position
angle as shown in Figure~\ref{fig:slit_locations}: LE325, LE326. Gaps in the
observed spectra correspond to areas contaminated by
sky-subtraction residuals.
\label{fig:4}}
\end{figure}

Observed LE spectra and corresponding dust-modeled isotropic spectra are
plotted as a function of PA in Figures~\ref{fig:6}-\ref{fig:4},
with each figure representing a field as shown
in Figure~\ref{fig:slit_locations}. Figure~\ref{fig:6} corresponds to the most
northern field. As previously mentioned, we should avoid searching for
differences between observed LE spectra. Instead, deviations in asymmetry are
defined by deviations from the
dust-modeled isotropic spectrum (red line) for each line of sight. A
deviation from the model then describes a deviation from the appropriate
weighted set of outburst spectra of SN~1987A as it was historically observed
along the direct line of sight.
\newpage
\subsubsection{\ha Profiles}
\label{sec:asymmetry:ha}
\begin{figure}
\epsscale{0.9}
\plotone{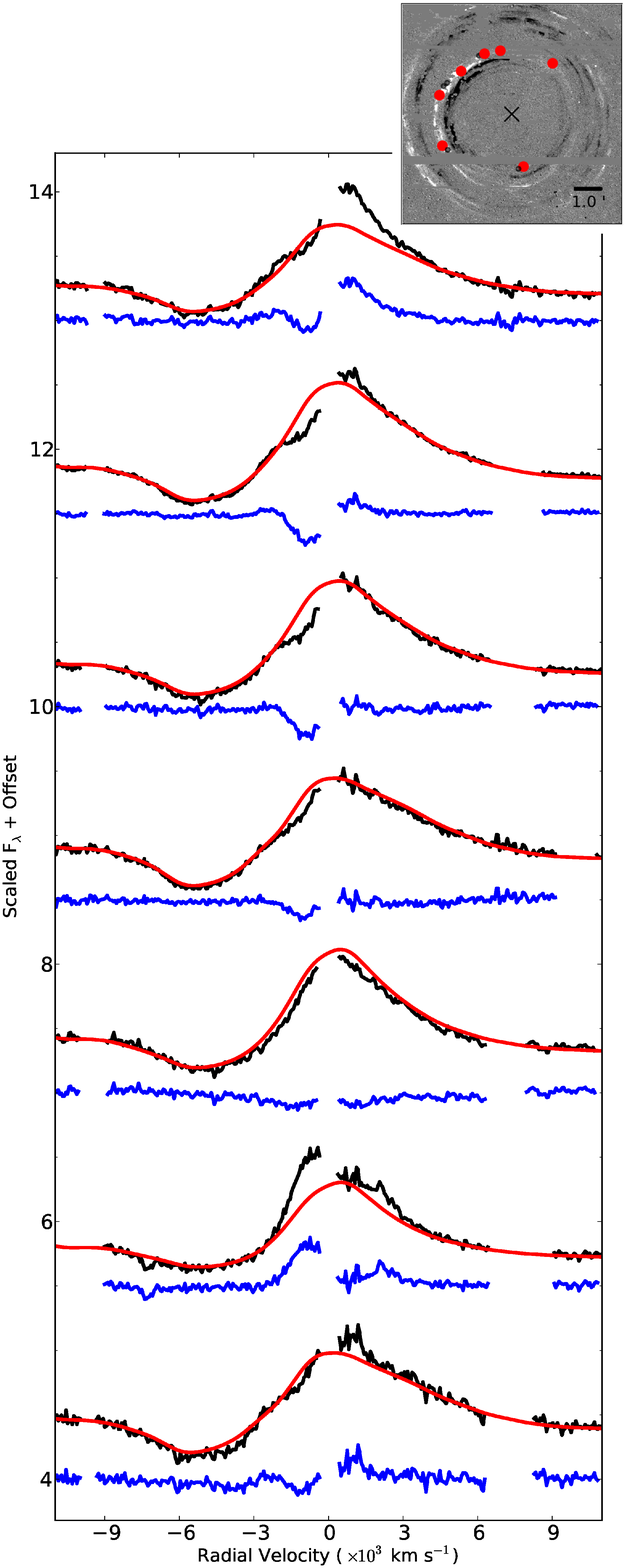}
\caption[]{Closeup of observed \ha profiles (black) and corresponding
dust-modeled isotropic spectra (red), as a function of PA for
seven viewing angles. The top \ha profile corresponds to the most northern
slit in the upper right legend, and the profiles increase in PA
from top to bottom: LE016, LE029, LE053, LE076, LE113, LE186, LE326. LE naming convention
 corresponds to PA of LE with respect to SNR. Residuals
between the observed LE spectra and the
dust-modeled isotropic spectra are plotted in blue below each profile. To
prevent guiding the eye, the portions of the \ha profile containing residuals
from sky subtraction have been masked out. The most northern profile, LE016,
has an excess in redshifted \ha emission and a blue knee at $\sim-2000$\kms
are observed. At almost opposite PA (LE0186, second from bottom), the
opposite asymmetry is observed in the \ha profile: an excess in blueshifted
emission and a red knee at $\sim+2200$\kms. Both asymmetry features diminish as
a function of viewing angle away from the north. Portions of the observed \ha
peak contaminated by sky-subtraction residuals have been removed to prevent
guiding the eye. No smoothing has been applied to any of the spectra.
\label{fig:allspecsha}}
\end{figure}
\begin{figure}
\epsscale{0.9}
\plotone{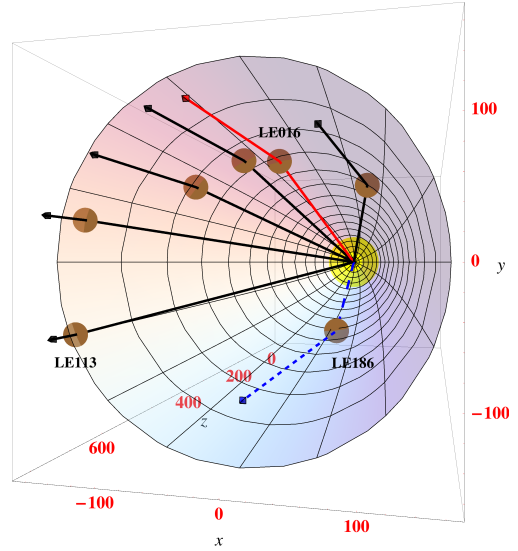}
\caption[]{3D locations of the scattering dust for the seven LEs from 
Figure~\ref{fig:allspecsha}. Highlighted in solid red and dashed blue are the 
extreme viewing angles corresponding to LE016 and LE186. North is towards 
the positive $y$ axis, east is towards the negative $x$ axis, and $z$ 
is the distance in front of the SN. All units are in light years.
\label{fig:3d}}
\end{figure}
\begin{figure}
\epsscale{0.9}
\plotone{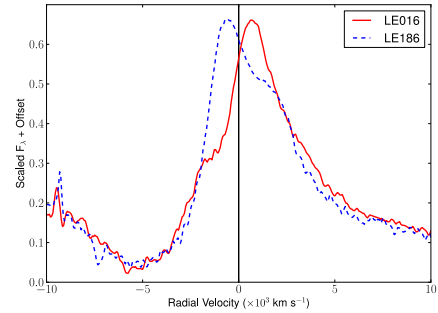}
\caption[]{Observed \ha lines from LE016 and LE186. Emission peaks have
been interpolated with high-order polynomials. Spectra are
scaled and offset for comparison purposes, as well as smoothed with a boxcar of
3 pixels. Although this plot does not take into account the
important differences in LE time-integrations between the spectra, it highlights
the overall difference in fine-structure in the two LE spectra from
opposite PAs. Observing \ha profiles with opposite asymmetry
structure at opposite PAs is surprising considering the opening
angle between the two LEs is $<40^{\circ}$. 
\label{fig:plain_cont_comparison}}
\end{figure}
\begin{figure}
\epsscale{0.9}
\plotone{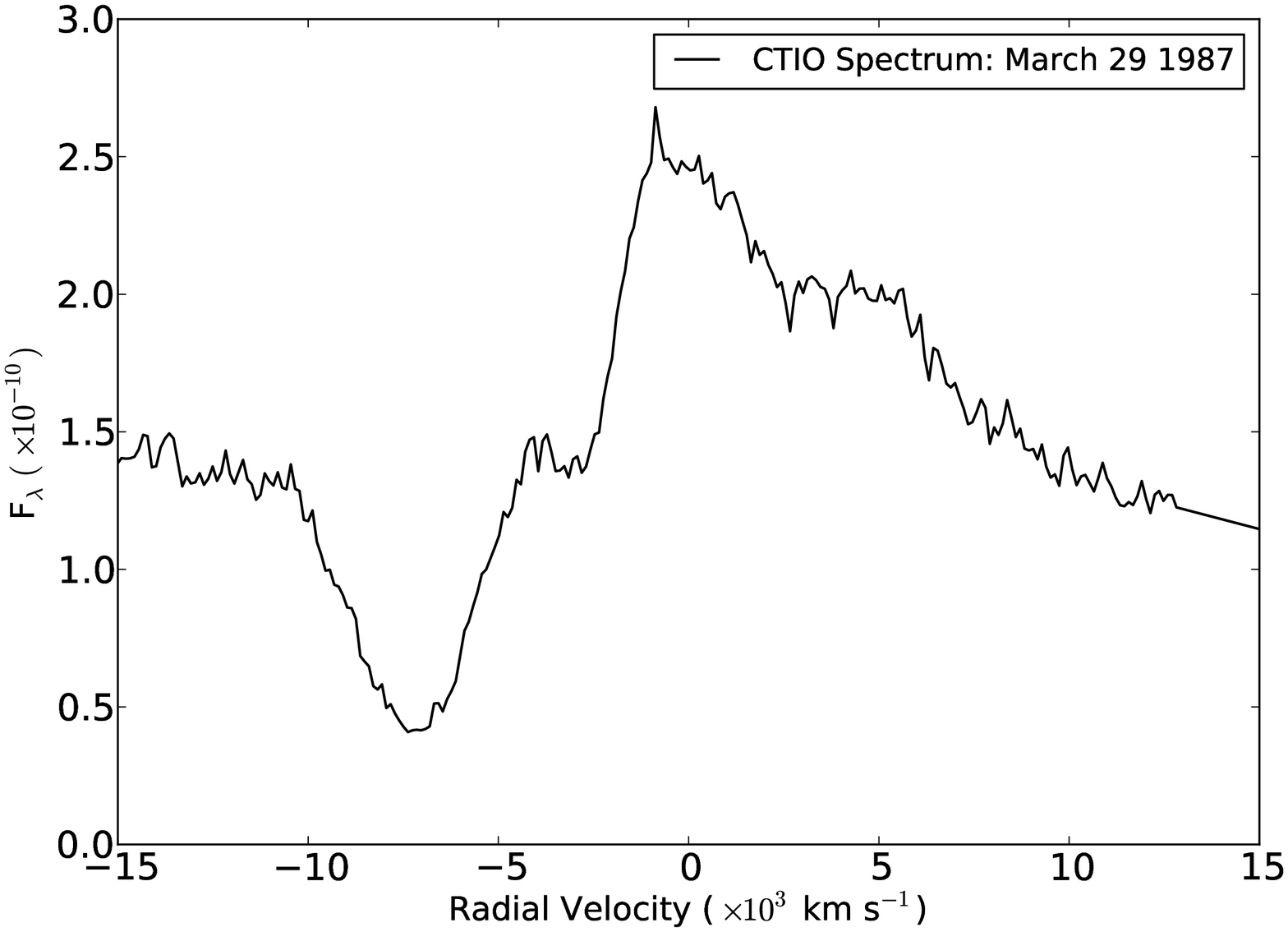}
\caption[]{Historical CTIO \ha profile from \citet{phillips88} of SN~1987A taken
on
March 29, 1987, 34 days after explosion. The fine-structure, showing blue and
red features at $\sim\pm3500$\kms, defines the ``Bochum
event'' which was observed 20-100 days after explosion. Evidence for this fine
structure is not present after the historical spectra are integrated with the
LE scattering model (red lines in Figure~\ref{fig:allspecsha}). \ha fine
structure observed in the LE spectra in Figure~\ref{fig:allspecsha} must
therefore require stronger line profile asymmetry than the original ``Bochum
event'' or the asymmetry must last for much longer than originally observed for
SN~1987A.
\label{fig:bochum}}
\end{figure}
Figures~\ref{fig:6}-\ref{fig:4} show that most features and line strengths
observed
in the optical LE spectra can be fit with the isotropic historical spectrum of
SN~1987A, without the need to invoke asymmetry in the outburst. However, the
fine-structure and strength of the \ha $\lambda6563$ line show deviations from
symmetry as a function of PA (i.e. viewing angle onto the SN).

Figure~\ref{fig:allspecsha} shows a closeup of the \ha profiles of seven unique
viewing angles as a function of increasing PA from top to
bottom (the geometry of the viewing angles is shown in Figure~\ref{fig:3d}). The most
northern line of sight LE, LE016, shows a strong blue knee in the profile at
$\sim-2000$\kms in addition to a strong excess of emission that is redshifted
from the rest wavelength by $\sim+800$\kms (and by $\sim+500$\kms compared to
the historical spectrum). The most southern LE, LE186, has a PA
almost directly opposite that of LE016. Its \ha profile (second from bottom in
Figure~\ref{fig:allspecsha}) shows a red knee in the fine-structure at
$\sim+2200$\kms and an excess in emission shifted towards the blue by
$\sim-500$\kms (and by $\sim-1000$\kms compared to the
historical spectrum). The qualitative difference in the \ha profile shape
between the two extreme viewing angles, LE016 and LE186, is shown in
Figure~\ref{fig:plain_cont_comparison}. This figure does not take into
account the effects of the LE observations on the integrated spectra and so
cannot be used as a direct comparison between the viewing angles. However, it
does qualitatively describe the asymmetry in the fine-structure
between the two viewing angles.

The two asymmetries in the \ha profile (summarized in
Figure~\ref{fig:plain_cont_comparison}) are the fine-structure, with a blue knee
in the northern profiles and a red knee in the southern profile, as well as
an excess of \ha emission that is redshifted in the north and blueshifted
in the south. Both of these asymmetries appear to be physical for a number
of reasons. The excess in \ha emission as well as the blue knee feature smoothly
diminish in Figure~\ref{fig:allspecsha} as PA is increased. At
PA $\sim110^{\circ}$ in the south-east quadrant, both the \ha
strength and fine-structure from LEs LE113 and LE117 can be fit with the
historical model spectrum of SN~1987A. The fact that both \ha asymmetries
(fine-structure and redshifted emission) smoothly diminish as a function of
viewing angle
from north to south in the eastern half of the LE ring is strong evidence that
these asymmetries are physical.

This fine-structure in the \ha profile is similar to
the ``Bochum event'' (named after the 61 cm Bochum telescope at La Silla, Chile) 
originally observed in SN~1987A
\citep{hanuschik87,phillips89}, where blue and red satellite emission features
were observed 20-100 days after explosion. Figure~\ref{fig:bochum} shows the \ha
profile of a CTIO spectrum taken 34 days after explosion \citep{phillips88}. The
blue and red emission satellites are similar to the blue and red
knees observed in the LE016 and LE186 LE spectra, respectively. It is tempting
to
compare the \ha fine-structure observed in Figure~\ref{fig:bochum} directly with
the fine-structure observed in the LE profiles. \emph{However, LE spectra
represent an
integration of many epochs of signal and cannot be compared directly to a
single epoch spectrum.} Although the ``Bochum event'' is prominent in
Figure~\ref{fig:bochum}, there is no evidence for fine-structure in any of the
isotropic model spectra in Figure~\ref{fig:allspecsha}. That is, the ``Bochum
event'' does not survive the smoothing effect when the historical spectra of
SN~1987A are integrated. This can be seen in the lower panel
of Figure~\ref{fig:modelsummary}, where the fine-structure from the ``Bochum
event'' is faintly visible in the maximum light spectrum, but entirely
nonexistent in the integrated model spectra. The fact that fine-structure
similar to the ``Bochum event'' is present in the observed LE spectra is
evidence for underlying
fine-structure that is much stronger than that originally observed for SN~1987A.
In general,
small deviations from symmetry observed in a LE spectrum must represent very
large deviations from symmetry in the underlying outburst in order to remain
after the integrating effect of the LE phenomenon. This is an important aspect
of all LE spectroscopy work that is rarely emphasized.

The excess emission in the north and south from profiles LE016 and
LE186, respectively, cannot be explained within uncertainties. To obtain the
largest
temporal coverage we used both CTIO and SAAO spectra when integrating the model
spectrum. There exists known systematic differences between the datasets of the
original observations from the two locations \citep{hamuy90}. Using the two data
sets individually in our analysis pipeline can lead to differences in
the \ha strength by $5\%-10\%$ in our models, but cannot account for the
$30\%-40\%$ excess observed in the LE016 and LE186 spectra. We also stress that
our analysis is based on relative flux comparisons only.
\begin{figure}
\epsscale{1.2}
\plotone{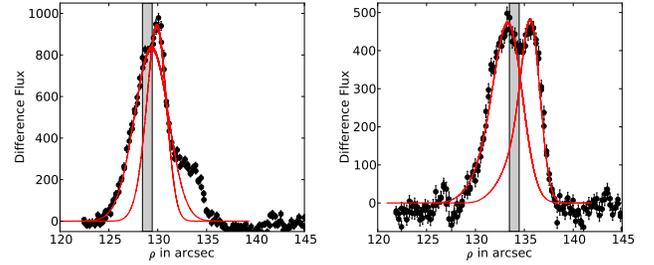}
\caption[]{LEFT: Observed LE profile on the sky for the southern LE186.
Over-plotted in red are models for two scattering dust
filaments which both contribute signal to the slit (plotted as the shaded
region). Although the narrow filament at larger $\rho$ contributes late-time
epochs to the spectrum, allowing for a larger emission-to-absorption ratio, the
excess emission in the LE186 \ha profile in Figure~\ref{fig:allspecsha} cannot
be modeled with the historical spectra of SN~1987A. See Appendix for full
consideration. RIGHT: Observed LE profile of LE180. See Appendix for full
consideration on why the LE fitting algorithm does not provide meaningful
results in this complex case.
\label{fig:dust_example_f3s32}}
\end{figure}

The LE profile on the sky for LE016 is very similar to that shown in
Figure~\ref{fig:modelsummary}. Since it is a single peak, the theoretical
maximum amount of \ha emission in the model would correspond to integrating the
historical spectra with the full lightcurve of SN~1987A out to $t=t_{now}$,
rather than an effective lightcurve that is truncated by a window function (i.e.
having an infinitely thick dust sheet). However, even this unphysical limit
cannot
reproduce the \ha emission-to-absorption ratio that is seen in
the
observed LE profile of LE016. The LE profile on the sky and the slit location
for
LE186, which has two closely spaced filaments, are shown in
Figure~\ref{fig:dust_example_f3s32}. Since the slit was placed on the first
peak,
there will be late-time nebular emission entering the slit from the second peak
at larger $\rho$. In such a case, it is possible to obtain a larger
emission-to-absorption ratio. However, we have taken this into account by
modeling both LE peaks and determining the relative contributions from each
peak entering the slits. This effect, discussed in more detail in the Appendix,
cannot account for the excess in emission
in LE186. We also stress that the apparent lack of any \ha emission excess in
LE180 is due to limitations in our LE fitting algorithrims when dealing with
such low signal-to-noise data. LE180 is discussed in more detail in the
Appendix. It should be noted, however, that LE180 appears to show the same
fine-structure in \ha (blueshifted emission peak and red knee) despite the low
signal of the spectrum.

The opening angles probed by the LE lines of sight are $<45\deg$. The fact
that large asymmetries are seen in the observed integrated LE spectra for
relatively small changes in viewing angle is evidence for a strong asymmetry in
the explosion.
\subsubsection{Additional Evidence of Asymmetry}
\label{sec:moreasymmetry}
Although the \ha profile shows the strongest asymmetry signature that is
dependent on viewing angle, we note here two additional possible sources of
asymmetry from the LE spectra: a velocity shift in the Fe II $\lambda5018$
line that appears to be correlated with the \ha asymmetry, and an unidentified
feature near $5265${\AA} present in only one direction.
\begin{figure}
\epsscale{1.2}
\plotone{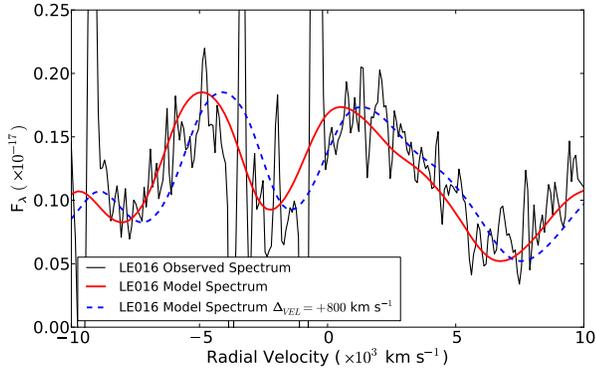}
\caption[]{Observed Fe II $\lambda5018$ line in LE016 most-northern LE
spectrum. The isotropic dust-modeled spectrum is over-plotted (solid red line),
along with the same isotropic spectrum shifted $800$\kms to the red (dashed blue
line), corresponding to the best fit to the peak location of the $\lambda5018$
line and of similar magnitude to the velocity shift observed in the \ha profile.
Note that the velocity-shifted
model spectrum is a better fit to the observed LE spectrum, indicating that the
\ha velocity asymmetry may be correlated with the velocity of the Fe II
$\lambda5018$ line.
\label{fig:f6s15_FeII_redshift}}
\end{figure}

Figure~\ref{fig:f6s15_FeII_redshift} plots the Fe II $\lambda5018$ line for the
observed LE spectra of LE016 (black), the asymmetric northern viewing angle.
The dust-modeled isotropic spectrum is plotted as solid red. We also plot in
dashed blue the same isotropic spectrum redshifted by $+800$\kms, which gives
the best fit to the emission peak of the Fe II $\lambda5018$ line (determined
by eye). The redshifted spectrum is a better fit to the line and is consistent
with the $600-800$\kms redshift observed in the \ha line for the same viewing
angle. As with the \ha line asymmetry, the magnitude of the
best-fitting redshift for the Fe II $\lambda5018$ line decreases as a function
of viewing angle away from the LE016 line of sight. For the case of LE186, the
line is weak and heavily contaminated from sky residuals. Although it has the
most blueward Fe II $\lambda5018$ line of all observed LE spectra, it is
unclear if the line is actually blueshifted with respect to the isotropic model
or to zero. The structure is complicated since this line is part of a
blend of the Fe II $\lambda\lambda4924$, $5018$, $5169$ features.
\begin{figure}
\epsscale{1.0}
\plotone{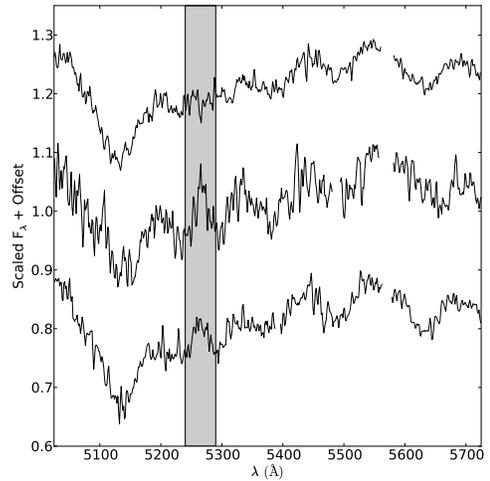}
\caption[]{Observed LE spectrum for LE016 (top) compared with the two
``equatorial'' LE locations (LE113
and LE117, middle and lower, respectively) shown at 
PA $\sim115\arcdeg$ in Figure~\ref{fig:slit_locations}. Both
spectra show an unidentified feature near $5265${\AA} that is not present in
any other viewing angle. The fact that the line appears broader in the upper
spectrum is consistent with that LE probing a narrower range
of early, high-velocity epochs compared to the lower spectrum as described in
Section~\ref{sec:mode_f2_example}.
\label{fig:f2_feature}}
\end{figure}

Figure~\ref{fig:f2_feature} shows an unidentified feature near $5265${\AA}
in the observed LE spectra for both ``equatorial'' (i.e. perpendicular to
maximum asymmetry axes $16$\arcdeg/$186$\arcdeg) lines of sight at
PA $\sim115\arcdeg$. Many LE spectra have
signal-to-noise ratios higher than the lower spectrum of
Figure~\ref{fig:f2_feature}, however there is no evidence for the feature in
any other line of sight. The fact that the feature is broader in
the LE117 (lower) spectrum is consistent with that LE probing a narrower range
of early, high-velocity epochs as described in
Section~\ref{sec:mode_f2_example}. Considering this wavelength region is
populated with many line blends and the fact that the spectra represent a
time-integration, it is currently unclear if the feature is due to differences
in chemical or velocity properties at this viewing angle.
\section{Discussion}
\label{sec:discussion}
Ignoring the smaller knee-like fine-structure, the \ha profiles of the LE016
and LE186 viewing angles in Figure~\ref{fig:plain_cont_comparison} argue 
strongly for a one-sided asymmetry in
SN~1987A. An overabundance of $^{56}$Ni in the southern far hemisphere would
create an excess in nonthermal excitation of hydrogen. This results in an excess
in redshifted emission for the northern LE016 viewing angle. If the
overabundance of $^{56}$Ni is inclined close to the plane of the sky (within
$21^{\circ}$), the overexcitation would occur in the near hemisphere with
respect to the LE186 line of sight, explaining the blueshifted \ha emission in
LE186. The question remains if the asymmetry summarized in 
Figure~\ref{fig:plain_cont_comparison} is strictly a result of this one-sided
asymmetry plus time-integration effects, or if an additional asymmetry is the
cause of the blue and red knee in the fine-structure of \ha in LE016 and LE186,
respectively.

The ``Bochum event'' (Figure~\ref{fig:bochum}) has previously
been interpreted as an asymmetric distribution of $^{56}$Ni 
\citep[e.g.,][]{lucy88,hanuschik90,chugai91a}.
Although both blue and red emission features are observed in
Figure~\ref{fig:bochum}, the blue emission feature was only additionally
observed
in H$\beta$ \citep{hanuschik90} while the redshifted emission feature was
also observed in infrared hydrogen lines \citep{larson87} and [Fe II] lines
\citep{haas90}. \citet{chugai91a} proposed a two-sided $^{56}$Ni asymmetry: (1)
a dominant $^{56}$Ni cloud in the far hemisphere producing the red emission
fine-structure and the redshifted emission peak at later epochs, and (2) a
smaller,
higher velocity
$^{56}$Ni cloud in the near hemisphere producing the blue fine-structure.
However, instead of invoking a smaller $^{56}$Ni cloud in the near hemisphere,
the more recent explanation for the blue fine-structure is a non-monotonic
Sobolev optical depth, $\tau(v)$, for \ha with a minimum near $v\approx5000\kms$
\citep{chugai91b,utrobin95,wang02,utrobin02,utrobin05}. This interpretation
does not require breaking spherical symmetry to explain the blue
fine-structure. Although \citet{utrobin05}
favor this scenario, their model cannot successfully produce the optical depth
minimum at the required strength. The smooth transition of the \ha profiles
in PA from LE016 to LE117 in Figure~\ref{fig:allspecsha} shows that the blue fine
structure is dependent on viewing angle. As such, \emph{some form} of deviation
from spherical symmetry must be the root of the blue fine-structure in the LE
\ha profiles and
presumably directly correlated to the blue fine-structure observed in the
``Bochum event.''

The ``Bochum event'' was observed on
days 20-100, while the redshift in lines of hydrogen and other elements was
observed after $\sim150$ days. For each LE spectrum,
Table~\ref{tab:modelresults} lists the approximate range of epochs that
contribute to the LE spectrum at greater than the $50\%$ level. This
relative contribution is based on the window function generated for each LE
(Figure~\ref{fig:modelsummary}), prior to any flux-weighted integration. While
the temporal resolution in the
LEs is not capable of distinguishing between photospheric and nebular epochs,
we can compare LEs with different degrees of nebular emission. It is therefore
interesting to note that \emph{both} the fine-structure and the velocity shift
of the \ha line appear to be more dominantly a function of viewing angle rather
than a function of epoch. LE066 probes the first $\sim80$ days of the explosion
only, but does not show fine-structure in \ha considerably different than the
nearby LEs probing much later epochs. The fact that the blue fine-structure
feature is also present in the early-epoch LE066 spectrum almost certainly
links this feature to an exaggerated version of the original ``Bochum event.''

Since the \ha P Cygni line is a blend with Ba II $\lambda6497$, it is possible
an asymmetry in the Ba II line is causing the observed blue fine-structure in
the LE spectra. However, \citet{utrobin95} determined that the
inclusion of the Ba II line was not sufficient to explain the blue emission in
the ``Bochum event.'' Additionally, unlike \ha, the Ba II $\lambda6142$ line
appears to be fit well with the isotropic model in the LE spectra.

A two-sided $^{56}$Ni model such as \citet{chugai91a} could explain the
\ha LE observations. A smaller high-velocity cloud blueshifted in the north
causing the blue and red fine-structure in the north and south viewing angles,
respectively. And a larger cloud redshifted in the south causing the red- and
blue-shifted excess in the emission observed in the north and south viewing
angles, respectively. The larger cloud could also be responsible
for the Fe II asymmetry shown in Figure~\ref{fig:f6s15_FeII_redshift}.
The original observations of SN~1987A provided strong evidence for the
large cloud being in the far hemisphere as previously stated. This is apparent
in the isotropic model spectrum of LE186, where the time-integrated emission
peak of \ha is shifted to the red. LE186 probes a larger range of epochs in the
explosion (out to $\sim400$ days), so the redshift and integrated line profile
asymmetry is most apparent in the LE186 isotropic model. The observed LE
spectrum, however, shows a small blueshift with respect to zero
velocity, implying the overabundance is inclined within $21^{\circ}$ of the
plane of the sky in the southern far hemisphere. If
the cloud was $\sim20^{\circ}$ from the plane of the sky, the radial velocities
would be a factor of $1.7-1.9$ larger in the north and south viewing
angles, more easily altering the \ha profile shape of the integrated LE
spectra.

Figure~\ref{fig:plain_cont_comparison} highlights 
the \ha emission peak redshifted and blueshifted with respect to zero 
velocity in the northern (LE016) and southern (LE186) viewing angles, 
respectively. With respect to the isotropic model of SN~1987A, 
the absolute velocity shift of the \ha peak is twice as large in the south 
($-1000$\kms in LE186 compared to $+500$\kms in LE016). This is 
surprising considering both LEs have nearly equal $\sim20^{\circ}$ 
north and south lines of sight onto the photosphere. However, LE186 
probes over twice the range of epochs in the original explosion 
compared to LE016. The velocity 
shift is most likely more apparent at these later nebular epochs as 
previously discussed, potentially accounting for the larger velocity 
shift for LE186. 

In addition to SN~1987A, a velocity shift in the \ha peak or a Bochum-like
event has also been observed for the Type~IIP SNe SN~1999em, SN~2000cb,
SN~2004dj,
and SN~2006ov, as well as the Type~II SN~2006bc
\citep{elmhamdi03,kleiser11,chugai05,chornock10,gallagher12}. The \ha profiles
of SN~2004dj are strikingly similar to that of the LE186 profile in
Figure~\ref{fig:plain_cont_comparison}. Although LE
spectra cannot be compared directly
with instantaneous spectra, it appears the \ha lines from \citet{chugai05} would
resemble that of LE186 after being smoothed by time-integration.
\citet{chugai05} were able to successfully model the SN~2004dj \ha
fine-structure using an asymmetric bipolar
$^{56}$Ni distribution, including a spherical component with two cylindrical
components. They modeled the observer viewing a dominant jet $30^{\circ}$
off-axis, resulting in a larger blueshifted emission peak with a smaller
redshifted feature in the profiles of \ha. This describes the LE186 LE
profile in Figure~\ref{fig:plain_cont_comparison}, suggesting a bipolar
$^{56}$Ni distribution as a possible explanation. Since at the opposite PA,
 LE016, we see the opposite set of features in
Figure~\ref{fig:plain_cont_comparison}, the two LE viewing angles would have to
be looking towards opposite ends of the bipolar distribution in the model of
\citet{chugai05}. Since the viewing angles only differ by $\sim40^{\circ}$,
this would put strict constraints on the orientation of the bipolar
distribution. This is, however, consistent with the dominant $^{56}$Ni
component being aligned within $\sim20^{\circ}$ to the plane of the sky, as
previously suggested.
\begin{figure}
\centering
\mbox{\subfigure{\includegraphics[height=1.55in]{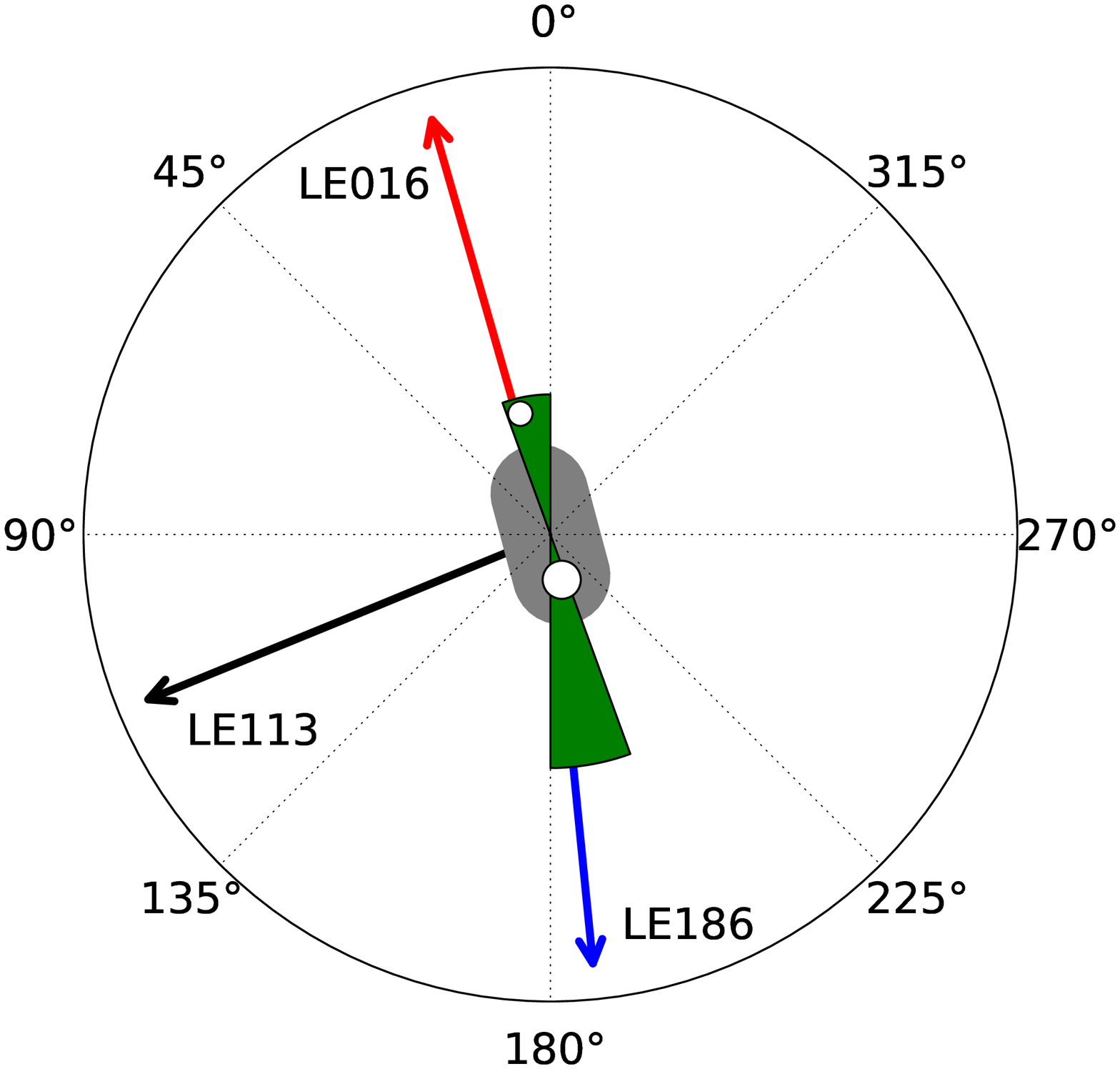}}\quad
\subfigure{\includegraphics[height=1.55in]{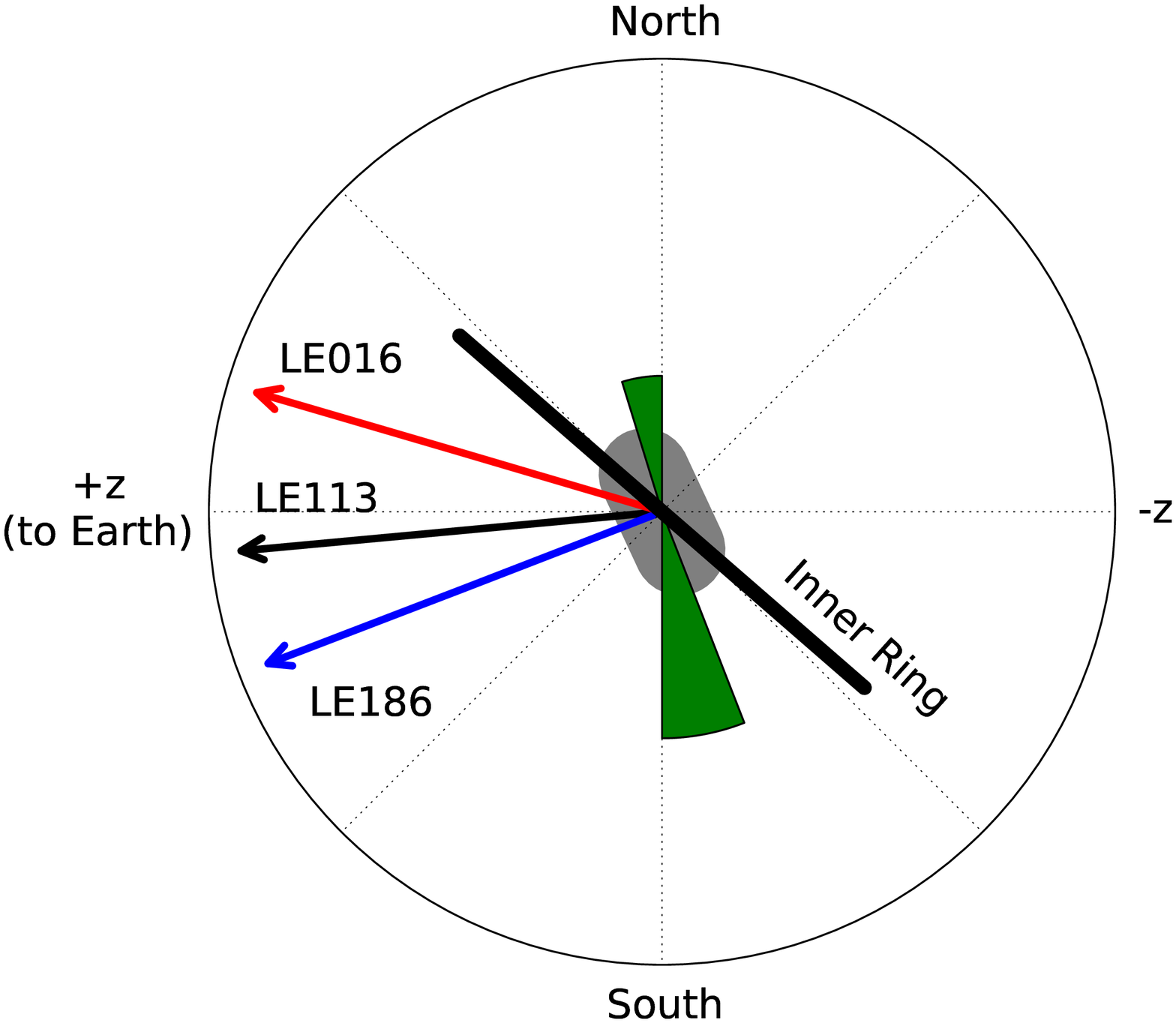} }}
\caption{LEFT: PAs on the sky of the three dominant LE viewing angles 
LE016, LE113 and LE186. The central grey region denotes the orientation 
on the sky of the elongated remnant ejecta (PA = $16^{\circ}$) from \citet{kjaer10}. 
The green wedges illustrate the proposed two-sided distribution of $^{56}$Ni, most 
dominant in the southern hemisphere. White circles illustrate the locations of the 
two mystery spots as identified by \citet{nisenson99}, with radius proportional to 
relative brightness in magnitudes of the two sources. Only the \emph{relative} distance 
from the center of the SNR to the mystery spots is to scale in the image. RIGHT: Schematic with viewing angle 
perpendicular to Earth's line of sight. The inclination of the inner circumstellar 
ring is shown along with the proposed two-sided distribution of $^{56}$Ni in 
green. Note that the green wedges are to highlight the proposed 
geometry (not absolute velocity) of the $^{56}$Ni asymmetry probed by the LE spectra, illustrating 
that the southern overabundance is most dominant.}
\label{fig:schematic}
\end{figure}

As noted in the introduction, one of the benefits of LE spectroscopy is it
allows the signatures
of the explosion in the first few hundred days to be compared directly with the
state of the remnant. The PA of the symmetry axis of the elongated ejecta
was measured to be $14^{\circ}\pm5^{\circ}$ using HST imaging \citep{wang02} and
$15^{\circ}\pm0.9^{\circ}$ using IFU spectra \citep{kjaer10}. This PA
corresponds to LE016, where we see the maximum deviation from symmetry in
\ha in the northern hemisphere. \citet{kjaer10} found the
present-day ejecta to be blueshifted in the north and redshifted in the south,
inclined out of the plane of the sky by $\sim25^{\circ}$. The bipolar $^{56}$Ni
distribution proposed above to explain the LE observations is therefore aligned
with the roughly $25$ year-old ejecta both in PA and inclination out of the
sky. The inner ring of circumstellar material is inclined $49^{\circ}$ out of
the plane of the sky, blueshifted in the north \citep{sugerman05}. The outer
circumstellar rings are similarly inclined, presumably related to the rotation
axis of the progenitor. The elongated ejecta and $^{56}$Ni distribution
probed by the LE observations are therefore aligned approximately in plane
with the equatorial ring, as opposed to sharing a symmetry axis as initially
proposed by \citet{wang02}. This disfavors the axially symmetric jet-induced
explosion model for SN~1987A proposed by \citet{wang02}. We illustrate 
the  proposed asymmetry with a schematic in Figure~\ref{fig:schematic}, 
highlighting the overabundance of $^{56}$Ni close to the plane of the sky 
and redshifted in the southern hemisphere.

The ``mystery spot'' of SN~1987A was a bright source observed in speckle
interferometry measurements 30, 38, and 50 days after the explosion, separated
by only 60 mas from the SN \citep{nisenson87,meikle87}, and appearing at 
PA=$194^{\circ}\pm2^{\circ}$. \citet{nisenson99} reprocessed this data,
and identified \emph{two} mystery spots in the data: (1) the original bright source 
at PA=$194^{\circ}\pm3^{\circ}$ separated by $60\pm8$~mas from the SNR, and 
(2) a fainter source at PA=$14^{\circ}\pm3^{\circ}$ separated by $160\pm8$~mas 
from the SNR. In order to be associated with the SN,
they claimed the mystery spots must be at relativistic speeds with the northern
spot blueshifted. A satisfactory explanation of the mystery spot (or
spots) has yet to appear, a fact which is often forgotten. Figure~\ref{fig:schematic} 
highlights the location of the mystery spots of \citet{nisenson99} with respect to our 
LEs. Since the PAs for the LE
\ha spectra showing maximum asymmetry ($16^{\circ}/186^{\circ}$) match the mystery spot PAs to within
$10^{\circ}$, future modeling of the LE lines may aid in an explanation of the mystery spots.

If in fact the blue fine-structure of the ``Bochum event'' is due to an
asymmetrical $^{56}$Ni feature in the northern hemisphere, as the LE spectra
here suggest, $^{56}$Ni is transported to even higher velocities than previously
considered. The blue feature emerges after 20 days at a radial velocity of
$-5000\kms$ in the \ha profile. Since the blue fine-structure is most dominant
in LE016 at a scattering angle of $17^{\circ}$, the lower limit of the absolute
velocity of the $^{56}$Ni cloud in the near hemisphere is $5200\kms$, a
velocity difficult to explain in current neutrino-driven explosion models.
Absolute velocities of $>7000\kms$ are required if the cloud is
inclined within $45^{\circ}$ to the plane of the sky and upwards of $10000\kms$
if within $30^{\circ}$. Assuming the dominant southern $^{56}$Ni overabundance
(and change in \ha peak velocity) is correlated to the red emission feature
observed in the ``Bochum event'' requires absolute velocities of $\sim10000\kms$
for that feature for an inclination of $20^{\circ}$ out of the plane of the
sky. For the case of SN~2000cb, \citet{utrobin11} required
radial mixing $^{56}$Ni to velocities of $8400\kms$ to reproduce the observed
lightcurve of SN~2000cb, although it was a more energetic Type IIP than
SN~1987A. Therefore, these unrealistically high velocities of
$^{56}$Ni for current neutrino-powered core-collapse explosion models require us
to step back and view the asymmetries in Figure~\ref{fig:plain_cont_comparison}
at their most basic level: a non-spherical excitation structure in the
early ejecta. \citet{dessart11} demonstrated that the P Cygni line
profiles from non-spherical Type II ejecta can be altered significantly
depending on viewing angle. Only future modeling of the LE spectra, taking into
account the effects of time-integration, will allow the ejecta geometry to be
determined or further constrained.

\section{Conclusions}
\label{sec:conclusions}
We have obtained optical spectra from the LE system of SN~1987A, allowing
time-integrated spectra of the first few hundred days of the original explosion
to be viewed from multiple lines of sight. We have modeled the LE spectra
using the original photometry and spectroscopy of SN~1987A using the model of
\citet{leprofile}. Using specific examples we have showed the model correctly
interprets LE flux profiles and spectra when both scattering dust properties and
observational properties are taken into consideration.

Each PA on the LE system represents a unique viewing angle onto the
photosphere which can be spectroscopically compared to the isotropic spectrum
calculated from the original outburst of SN~1987A. The LE spectra show
evidence for asymmetry, demonstrating the technique of targeted LE
spectroscopy as a useful probe for observing SN asymmetries. The
observed asymmetries can be summarized as follows:
\begin{enumerate}
 \item Fine-structure in the \ha line stronger than the original ``Bochum
event'' is observed most strongly at PAs $16^{\circ}$ and $186^{\circ}$, with
the \ha profiles showing opposite asymmetry features in the north and south
viewing angles.
 \item At PA $16^{\circ}$ we observed an excess in redshifted \ha emission and a
blueshifted knee. At PA $186^{\circ}$ we observed an excess of blueshifted \ha
emission and a red knee. This fine-structure diminishes slowly as the PA increases
from $16^{\circ}$ to $16^{\circ}+90^{\circ}$, with the LEs perpendicular to the
symmetry axis showing no evidence for asymmetry.
 \item At PA $16^{\circ}$ we observe a velocity shift in the Fe II $\lambda5018$
line of the same magnitude and direction as the velocity shift in the \ha
emission peak at the same viewing angle. As with the \ha fine-structure, this
velocity shift appears to diminish as PA increases from $16^{\circ}$.
 \item At PA $\sim115^{\circ}$ (roughly perpendicular to symmetry axis defined
by $16^{\circ}/186^{\circ}$ viewing angles) we observed an unidentified feature
near $\lambda5265${\AA} not observed in any other viewing angle.
\end{enumerate}

This symmetry axis defined by the $16^{\circ}/186^{\circ}$ viewing angles is in
excellent agreement with the current axis of symmetry in the ejecta geometry,
the initial polarization and speckle observations, as well as the location
of the ``mystery spot.'' The \ha lines at PAs $16^{\circ}$ and $186^{\circ}$
are very similar to the \ha lines observed in SN~2004dj and modeled as a
two-sided $^{56}$Ni distribution by \citet{chugai05}. This same model could
describe a two-sided ejection of $^{56}$Ni in SN~1987A as probed by the LE \ha
lines. The $^{56}$Ni is blueshifted in the north and redshifted in the south,
with the dominant overabundance of $^{56}$Ni being inclined $\sim20^{\circ}$
from the plane of the sky. The indication that high-velocity $^{56}$Ni is not 
correlated with the inner ring and presumed rotation axis may indicate that the 
explosion mechanism is independent of rotation. While these observations argue for a two-sided distribution of high-velocity 
$^{56}$Ni, at their most basic level they probe unequal source functions of the \ha 
line at different viewing angles. Only future modeling of the LE spectra will be able 
to constrain the early ejecta geometry with confidence.
\acknowledgements
We thank John Menzies for organizing and providing the original SAAO spectra of
SN~1987A. We also thank the anonymous referee for very useful comments. BPS
thanks Rollin Thomas and Tomasz Plewa for helpful discussion. DLW
acknowledges support from the Natural Sciences and Engineering
Research Council of Canada (NSERC). SuperMACHO was supported by the HST
grant GO-10583 and GO-10903.
\bibliographystyle{apj}
\bibliography{sinnott2012_87a}

\begin{thebibliography}{71}
\expandafter\ifx\csname natexlab\endcsname\relax\def\natexlab#1{#1}\fi

\bibitem[{{Arnett} {et~al.}(1989){Arnett}, {Bahcall}, {Kirshner}, \&
  {Woosley}}]{arnett89}
{Arnett}, W.~D., {Bahcall}, J.~N., {Kirshner}, R.~P., \& {Woosley}, S.~E. 1989,
  \araa, 27, 629

\bibitem[{{Bailey}(1988)}]{bailey88}
{Bailey}, J. 1988, Proceedings of the Astronomical Society of Australia, 7, 405

\bibitem[{{Blondin} {et~al.}(2003){Blondin}, {Mezzacappa}, \&
  {DeMarino}}]{blondin03}
{Blondin}, J.~M., {Mezzacappa}, A., \& {DeMarino}, C. 2003, \apj, 584, 971

\bibitem[{{Burrows}(2012)}]{burrows12}
{Burrows}, A. 2012, ArXiv e-prints

\bibitem[{{Caldwell} {et~al.}(1993){Caldwell}, {Menzies}, {Banfield},
  {Catchpole}, {Whitelock}, {Feast}, {Lloyd Evans}, {Sekiguchi}, {Zijlstra},
  {Allen}, {Bell}, {Blades}, {Buckley}, {Byrne}, {Callanan}, {Collins},
  {Cumming}, {O'Donoghue}, {Fairall}, {Freeman}, {Holmgren}, {Jones}, {Latham},
  {Maddox}, {Meadows}, {Meikle}, {Mittaz}, {Monk}, {Penny}, {Pollacco},
  {Slawson}, {Soltynski}, {Spyromilio}, {Stirpe}, {Stobie}, \&
  {Willmer}}]{87aspectra7}
{Caldwell}, J.~A.~R., {et~al.} 1993, \mnras, 262, 313

\bibitem[{{Catchpole} {et~al.}(1987){Catchpole}, {Menzies}, {Monk}, {Wargau},
  {Pollaco}, {Carter}, {Whitelock}, {Marang}, {Laney}, {Balona}, {Feast},
  {Lloyd Evans}, {Sekiguchi}, {Laing}, {Kilkenny}, {Spencer Jones}, {Roberts},
  {Cousins}, {van Vuuren}, \& {Winkler}}]{87aspectra2}
{Catchpole}, R.~M., {et~al.} 1987, \mnras, 229, 15P

\bibitem[{{Catchpole} {et~al.}(1988){Catchpole}, {Whitelock}, {Feast},
  {Menzies}, {Glass}, {Marang}, {Laing}, {Spencer Jones}, {Roberts}, {Balona},
  {Carter}, {Laney}, {Evans}, {Sekiguchi}, {Hutchinson}, {Maddison},
  {Albinson}, {Evans}, {Allen}, {Winkler}, {Fairall}, {Corbally}, {Davies}, \&
  {Parker}}]{87aspectra3}
---. 1988, \mnras, 231, 75P

\bibitem[{{Catchpole} {et~al.}(1989){Catchpole}, {Whitelock}, {Menzies},
  {Feast}, {Marang}, {Sekiguchi}, {van Wyk}, {Roberts}, {Balona}, {Egan},
  {Carter}, {Laney}, {Laing}, {Spencer Jones}, {Glass}, {Winkler}, {Fairall},
  {Lloyd Evans}, {Cropper}, {Shenton}, {Hill}, {Payne}, {Jones}, {Wargau},
  {Mason}, {Jeffery}, {Hellier}, {Parker}, {Chini}, {James}, {Doyle}, {Butler},
  \& {Bromage}}]{87aspectra5}
---. 1989, \mnras, 237, 55P

\bibitem[{{Chornock} {et~al.}(2010){Chornock}, {Filippenko}, {Li}, \&
  {Silverman}}]{chornock10}
{Chornock}, R., {Filippenko}, A.~V., {Li}, W., \& {Silverman}, J.~M. 2010,
  \apj, 713, 1363

\bibitem[{{Chugai}(1991{\natexlab{a}})}]{chugai91a}
{Chugai}, N.~N. 1991{\natexlab{a}}, \sovast, 35, 171

\bibitem[{{Chugai}(1991{\natexlab{b}})}]{chugai91b}
---. 1991{\natexlab{b}}, Soviet Astronomy Letters, 17, 400

\bibitem[{{Chugai} {et~al.}(2005){Chugai}, {Fabrika}, {Sholukhova},
  {Goranskij}, {Abolmasov}, \& {Vlasyuk}}]{chugai05}
{Chugai}, N.~N., {Fabrika}, S.~N., {Sholukhova}, O.~N., {Goranskij}, V.~P.,
  {Abolmasov}, P.~K., \& {Vlasyuk}, V.~V. 2005, Astronomy Letters, 31, 792

\bibitem[{{Cropper} {et~al.}(1988){Cropper}, {Bailey}, {McCowage}, {Cannon}, \&
  {Couch}}]{cropper88}
{Cropper}, M., {Bailey}, J., {McCowage}, J., {Cannon}, R.~D., \& {Couch}, W.~J.
  1988, \mnras, 231, 695

\bibitem[{{Crotts}(1988)}]{crotts88}
{Crotts}, A.~P.~S. 1988, \apjl, 333, L51

\bibitem[{{Dessart} \& {Hillier}(2011)}]{dessart11}
{Dessart}, L., \& {Hillier}, D.~J. 2011, \mnras, 415, 3497

\bibitem[{{Elmhamdi} {et~al.}(2003){Elmhamdi}, {Danziger}, {Chugai},
  {Pastorello}, {Turatto}, {Cappellaro}, {Altavilla}, {Benetti}, {Patat}, \&
  {Salvo}}]{elmhamdi03}
{Elmhamdi}, A., {et~al.} 2003, \mnras, 338, 939

\bibitem[{{Gallagher} {et~al.}(2012){Gallagher}, {Sugerman}, {Clayton},
  {Andrews}, {Clem}, {Barlow}, {Ercolano}, {Fabbri}, {Otsuka}, {Wesson}, \&
  {Meixner}}]{gallagher12}
{Gallagher}, J.~S., {et~al.} 2012, \apj, 753, 109

\bibitem[{{Garg} {et~al.}(2007){Garg}, {Stubbs}, {Challis}, {Wood-Vasey},
  {Blondin}, {Huber}, {Cook}, {Nikolaev}, {Rest}, {Smith}, {Olsen}, {Suntzeff},
  {Aguilera}, {Prieto}, {Becker}, {Miceli}, {Miknaitis}, {Clocchiatti},
  {Minniti}, {Morelli}, \& {Welch}}]{garg07}
{Garg}, A., {et~al.} 2007, \aj, 133, 403

\bibitem[{{Gawryszczak} {et~al.}(2010){Gawryszczak}, {Guzman}, {Plewa}, \&
  {Kifonidis}}]{gawryszczak10}
{Gawryszczak}, A., {Guzman}, J., {Plewa}, T., \& {Kifonidis}, K. 2010, \aap,
  521, A38

\bibitem[{{Gouiffes} {et~al.}(1988){Gouiffes}, {Rosa}, {Melnick}, {Danziger},
  {Remy}, {Santini}, {Sauvageot}, {Jakobsen}, \& {Ruiz}}]{gouiffes88}
{Gouiffes}, C., {et~al.} 1988, \aap, 198, L9

\bibitem[{{Haas} {et~al.}(1990){Haas}, {Erickson}, {Lord}, {Hollenbach},
  {Colgan}, \& {Burton}}]{haas90}
{Haas}, M.~R., {Erickson}, E.~F., {Lord}, S.~D., {Hollenbach}, D.~J., {Colgan},
  S.~W.~J., \& {Burton}, M.~G. 1990, \apj, 360, 257

\bibitem[{{Hammer} {et~al.}(2010){Hammer}, {Janka}, \& {M{\"u}ller}}]{hammer10}
{Hammer}, N.~J., {Janka}, H.-T., \& {M{\"u}ller}, E. 2010, \apj, 714, 1371

\bibitem[{{Hamuy} \& {Suntzeff}(1990)}]{87aphoto_hamuy90}
{Hamuy}, M., \& {Suntzeff}, N.~B. 1990, \aj, 99, 1146

\bibitem[{{Hamuy} {et~al.}(1990){Hamuy}, {Suntzeff}, {Bravo}, \&
  {Phillips}}]{hamuy90}
{Hamuy}, M., {Suntzeff}, N.~B., {Bravo}, J., \& {Phillips}, M.~M. 1990, \pasp,
  102, 888

\bibitem[{{Hamuy} {et~al.}(1988){Hamuy}, {Suntzeff}, {Gonzalez}, \&
  {Martin}}]{87aphoto_hamuy88}
{Hamuy}, M., {Suntzeff}, N.~B., {Gonzalez}, R., \& {Martin}, G. 1988, \aj, 95,
  63

\bibitem[{{Hanuschik} \& {Dachs}(1987)}]{hanuschik87}
{Hanuschik}, R.~W., \& {Dachs}, J. 1987, \aap, 182, L29

\bibitem[{{Hanuschik} \& {Thimm}(1990)}]{hanuschik90}
{Hanuschik}, R.~W., \& {Thimm}, G.~J. 1990, \aap, 231, 77

\bibitem[{{Jeffery}(1987)}]{jeffery87}
{Jeffery}, D.~J. 1987, \nat, 329, 419

\bibitem[{{Kim} {et~al.}(2008){Kim}, {Rieke}, {Krause}, {Misselt},
  {Indebetouw}, \& {Johnson}}]{kim08}
{Kim}, Y., {Rieke}, G.~H., {Krause}, O., {Misselt}, K., {Indebetouw}, R., \&
  {Johnson}, K.~E. 2008, \apj, 678, 287

\bibitem[{{Kj{\ae}r} {et~al.}(2010){Kj{\ae}r}, {Leibundgut}, {Fransson},
  {Jerkstrand}, \& {Spyromilio}}]{kjaer10}
{Kj{\ae}r}, K., {Leibundgut}, B., {Fransson}, C., {Jerkstrand}, A., \&
  {Spyromilio}, J. 2010, \aap, 517, A51

\bibitem[{{Kleiser} {et~al.}(2011){Kleiser}, {Poznanski}, {Kasen}, {Young},
  {Chornock}, {Filippenko}, {Challis}, {Ganeshalingam}, {Kirshner}, {Li},
  {Matheson}, {Nugent}, \& {Silverman}}]{kleiser11}
{Kleiser}, I.~K.~W., {et~al.} 2011, \mnras, 415, 372

\bibitem[{{Krause} {et~al.}(2008{\natexlab{a}}){Krause}, {Birkmann}, {Usuda},
  {Hattori}, {Goto}, {Rieke}, \& {Misselt}}]{krause08casa}
{Krause}, O., {Birkmann}, S.~M., {Usuda}, T., {Hattori}, T., {Goto}, M.,
  {Rieke}, G.~H., \& {Misselt}, K.~A. 2008{\natexlab{a}}, Science, 320, 1195

\bibitem[{{Krause} {et~al.}(2008{\natexlab{b}}){Krause}, {Tanaka}, {Usuda},
  {Hattori}, {Goto}, {Birkmann}, \& {Nomoto}}]{krause08tycho}
{Krause}, O., {Tanaka}, M., {Usuda}, T., {Hattori}, T., {Goto}, M., {Birkmann},
  S., \& {Nomoto}, K. 2008{\natexlab{b}}, \nat, 456, 617

\bibitem[{{Larson} {et~al.}(1987){Larson}, {Drapatz}, {Mumma}, \&
  {Weaver}}]{larson87}
{Larson}, H.~P., {Drapatz}, S., {Mumma}, M.~J., \& {Weaver}, H.~A. 1987, in
  European Southern Observatory Conference and Workshop Proceedings, Vol.~26,
  European Southern Observatory Conference and Workshop Proceedings, ed. I.~J.
  {Danziger}, 147--151

\bibitem[{{Lucy}(1988)}]{lucy88}
{Lucy}, L.~B. 1988, in Supernova 1987A in the Large Magellanic Cloud, ed.
  M.~{Kafatos} \& A.~G. {Michalitsianos}, 323--334

\bibitem[{{Meaburn} {et~al.}(1995){Meaburn}, {Bryce}, \&
  {Holloway}}]{meaburn95}
{Meaburn}, J., {Bryce}, M., \& {Holloway}, A.~J. 1995, \aap, 299, L1

\bibitem[{{Meikle} {et~al.}(1987){Meikle}, {Matcher}, \& {Morgan}}]{meikle87}
{Meikle}, W.~P.~S., {Matcher}, S.~J., \& {Morgan}, B.~L. 1987, \nat, 329, 608

\bibitem[{{Menzies} {et~al.}(1987){Menzies}, {Catchpole}, {van Vuuren},
  {Winkler}, {Laney}, {Whitelock}, {Cousins}, {Carter}, {Marang}, {Lloyd
  Evans}, {Roberts}, {Kilkenny}, {Spencer Jones}, {Sekiguchi}, {Fairall}, \&
  {Wolstencroft}}]{87aspectra1}
{Menzies}, J.~W., {et~al.} 1987, \mnras, 227, 39P

\bibitem[{{Miknaitis} {et~al.}(2007){Miknaitis}, {Pignata}, {Rest},
  {Wood-Vasey}, {Blondin}, {Challis}, {Smith}, {Stubbs}, {Suntzeff}, {Foley},
  {Matheson}, {Tonry}, {Aguilera}, {Blackman}, {Becker}, {Clocchiatti},
  {Covarrubias}, {Davis}, {Filippenko}, {Garg}, {Garnavich}, {Hicken}, {Jha},
  {Krisciunas}, {Kirshner}, {Leibundgut}, {Li}, {Miceli}, {Narayan}, {Prieto},
  {Riess}, {Salvo}, {Schmidt}, {Sollerman}, {Spyromilio}, \&
  {Zenteno}}]{miknaitis07}
{Miknaitis}, G., {et~al.} 2007, \apj, 666, 674

\bibitem[{{M{\"u}ller} {et~al.}(2012){M{\"u}ller}, {Janka}, \&
  {Marek}}]{muller12}
{M{\"u}ller}, B., {Janka}, H.-T., \& {Marek}, A. 2012, \apj, 756, 84

\bibitem[{{Nisenson} \& {Papaliolios}(1999)}]{nisenson99}
{Nisenson}, P., \& {Papaliolios}, C. 1999, \apjl, 518, L29

\bibitem[{{Nisenson} {et~al.}(1987){Nisenson}, {Papaliolios}, {Karovska}, \&
  {Noyes}}]{nisenson87}
{Nisenson}, P., {Papaliolios}, C., {Karovska}, M., \& {Noyes}, R. 1987, \apjl,
  320, L15

\bibitem[{{Papaliolios} {et~al.}(1989){Papaliolios}, {Krasovska}, {Koechlin},
  {Nisenson}, \& {Standley}}]{papaliolios89}
{Papaliolios}, C., {Krasovska}, M., {Koechlin}, L., {Nisenson}, P., \&
  {Standley}, C. 1989, \nat, 338, 565

\bibitem[{{Phillips} {et~al.}(1990){Phillips}, {Hamuy}, {Heathcote},
  {Suntzeff}, \& {Kirhakos}}]{phillips90}
{Phillips}, M.~M., {Hamuy}, M., {Heathcote}, S.~R., {Suntzeff}, N.~B., \&
  {Kirhakos}, S. 1990, \aj, 99, 1133

\bibitem[{{Phillips} \& {Heathcote}(1989)}]{phillips89}
{Phillips}, M.~M., \& {Heathcote}, S.~R. 1989, \pasp, 101, 137

\bibitem[{{Phillips} {et~al.}(1988){Phillips}, {Heathcote}, {Hamuy}, \&
  {Navarrete}}]{phillips88}
{Phillips}, M.~M., {Heathcote}, S.~R., {Hamuy}, M., \& {Navarrete}, M. 1988,
  \aj, 95, 1087

\bibitem[{{Rest} {et~al.}(2012{\natexlab{a}}){Rest}, {Sinnott}, \&
  {Welch}}]{lereview}
{Rest}, A., {Sinnott}, B., \& {Welch}, D.~L. 2012{\natexlab{a}}, Proceedings of
  the Astronomical Society of Australia, 29, 466

\bibitem[{{Rest} {et~al.}(2011{\natexlab{a}}){Rest}, {Sinnott}, {Welch},
  {Foley}, {Narayan}, {Mandel}, {Huber}, \& {Blondin}}]{leprofile}
{Rest}, A., {Sinnott}, B., {Welch}, D.~L., {Foley}, R.~J., {Narayan}, G.,
  {Mandel}, K., {Huber}, M.~E., \& {Blondin}, S. 2011{\natexlab{a}}, \apj, 732,
  2

\bibitem[{{Rest} {et~al.}(2005{\natexlab{a}}){Rest}, {Suntzeff}, {Olsen},
  {Prieto}, {Smith}, {Welch}, {Becker}, {Bergmann}, {Clocchiatti}, {Cook},
  {Garg}, {Huber}, {Miknaitis}, {Minniti}, {Nikolaev}, \& {Stubbs}}]{rest05}
{Rest}, A., {et~al.} 2005{\natexlab{a}}, \nat, 438, 1132

\bibitem[{{Rest} {et~al.}(2005{\natexlab{b}}){Rest}, {Stubbs}, {Becker},
  {Miknaitis}, {Miceli}, {Covarrubias}, {Hawley}, {Smith}, {Suntzeff}, {Olsen},
  {Prieto}, {Hiriart}, {Welch}, {Cook}, {Nikolaev}, {Huber}, {Prochtor},
  {Clocchiatti}, {Minniti}, {Garg}, {Challis}, {Keller}, \&
  {Schmidt}}]{supermacho}
---. 2005{\natexlab{b}}, \apj, 634, 1103

\bibitem[{{Rest} {et~al.}(2008{\natexlab{a}}){Rest}, {Welch}, {Suntzeff},
  {Oaster}, {Lanning}, {Olsen}, {Smith}, {Becker}, {Bergmann}, {Challis},
  {Clocchiatti}, {Cook}, {Damke}, {Garg}, {Huber}, {Matheson}, {Minniti},
  {Prieto}, \& {Wood-Vasey}}]{rest08a}
---. 2008{\natexlab{a}}, \apjl, 681, L81

\bibitem[{{Rest} {et~al.}(2008{\natexlab{b}}){Rest}, {Matheson}, {Blondin},
  {Bergmann}, {Welch}, {Suntzeff}, {Smith}, {Olsen}, {Prieto}, {Garg},
  {Challis}, {Stubbs}, {Hicken}, {Modjaz}, {Wood-Vasey}, {Zenteno}, {Damke},
  {Newman}, {Huber}, {Cook}, {Nikolaev}, {Becker}, {Miceli}, {Covarrubias},
  {Morelli}, {Pignata}, {Clocchiatti}, {Minniti}, \& {Foley}}]{rest08b}
---. 2008{\natexlab{b}}, \apj, 680, 1137

\bibitem[{{Rest} {et~al.}(2011{\natexlab{b}}){Rest}, {Foley}, {Sinnott},
  {Welch}, {Badenes}, {Filippenko}, {Bergmann}, {Bhatti}, {Blondin}, {Challis},
  {Damke}, {Finley}, {Huber}, {Kasen}, {Kirshner}, {Matheson}, {Mazzali},
  {Minniti}, {Nakajima}, {Narayan}, {Olsen}, {Sauer}, {Smith}, \&
  {Suntzeff}}]{casaspec}
---. 2011{\natexlab{b}}, \apj, 732, 3

\bibitem[{{Rest} {et~al.}(2012{\natexlab{b}}){Rest}, {Prieto}, {Walborn},
  {Smith}, {Bianco}, {Chornock}, {Welch}, {Howell}, {Huber}, {Foley}, {Fong},
  {Sinnott}, {Bond}, {Smith}, {Toledo}, {Minniti}, \& {Mandel}}]{eta}
---. 2012{\natexlab{b}}, \nat, 482, 375

\bibitem[{{Shigeyama} \& {Nomoto}(1990)}]{shigeyama90}
{Shigeyama}, T., \& {Nomoto}, K. 1990, \apj, 360, 242

\bibitem[{{Spyromilio} {et~al.}(1990){Spyromilio}, {Meikle}, \&
  {Allen}}]{spyromilio90}
{Spyromilio}, J., {Meikle}, W.~P.~S., \& {Allen}, D.~A. 1990, \mnras, 242, 669

\bibitem[{{Sugerman} {et~al.}(2005){Sugerman}, {Crotts}, {Kunkel}, {Heathcote},
  \& {Lawrence}}]{sugerman05}
{Sugerman}, B.~E.~K., {Crotts}, A.~P.~S., {Kunkel}, W.~E., {Heathcote}, S.~R.,
  \& {Lawrence}, S.~S. 2005, \apjs, 159, 60

\bibitem[{{Suntzeff} {et~al.}(1988{\natexlab{a}}){Suntzeff}, {Hamuy}, {Martin},
  {Gomez}, \& {Gonzalez}}]{87aphoto_suntzeff88}
{Suntzeff}, N.~B., {Hamuy}, M., {Martin}, G., {Gomez}, A., \& {Gonzalez}, R.
  1988{\natexlab{a}}, \aj, 96, 1864

\bibitem[{{Suntzeff} {et~al.}(1988{\natexlab{b}}){Suntzeff}, {Heathcote},
  {Weller}, {Caldwell}, \& {Huchra}}]{suntzeff88}
{Suntzeff}, N.~B., {Heathcote}, S., {Weller}, W.~G., {Caldwell}, N., \&
  {Huchra}, J.~P. 1988{\natexlab{b}}, \nat, 334, 135

\bibitem[{{Utrobin}(2004)}]{utrobin04}
{Utrobin}, V.~P. 2004, Astronomy Letters, 30, 293

\bibitem[{{Utrobin} \& {Chugai}(2002)}]{utrobin02}
{Utrobin}, V.~P., \& {Chugai}, N.~N. 2002, Astronomy Letters, 28, 386

\bibitem[{{Utrobin} \& {Chugai}(2005)}]{utrobin05}
---. 2005, \aap, 441, 271

\bibitem[{{Utrobin} \& {Chugai}(2011)}]{utrobin11}
---. 2011, \aap, 532, A100

\bibitem[{{Utrobin} {et~al.}(1995){Utrobin}, {Chugai}, \&
  {Andronova}}]{utrobin95}
{Utrobin}, V.~P., {Chugai}, N.~N., \& {Andronova}, A.~A. 1995, \aap, 295, 129

\bibitem[{{van Dokkum}(2001)}]{vandokkum01}
{van Dokkum}, P.~G. 2001, \pasp, 113, 1420

\bibitem[{{Wang} \& {Wheeler}(2008)}]{wang08}
{Wang}, L., \& {Wheeler}, J.~C. 2008, \araa, 46, 433

\bibitem[{{Wang} {et~al.}(2002){Wang}, {Wheeler}, {H{\"o}flich}, {Khokhlov},
  {Baade}, {Branch}, {Challis}, {Filippenko}, {Fransson}, {Garnavich},
  {Kirshner}, {Lundqvist}, {McCray}, {Panagia}, {Pun}, {Phillips}, {Sonneborn},
  \& {Suntzeff}}]{wang02}
{Wang}, L., {et~al.} 2002, \apj, 579, 671

\bibitem[{{Weingartner} \& {Draine}(2001)}]{weingartner01}
{Weingartner}, J.~C., \& {Draine}, B.~T. 2001, \apj, 548, 296

\bibitem[{{Whitelock} {et~al.}(1988){Whitelock}, {Catchpole}, {Menziez},
  {Feast}, {Winkler}, {Marang}, {Glass}, {Balona}, {Egan}, {Carter}, {Roberts},
  {Sekiguchi}, {Laney}, {Lloyd Evans}, {Laing}, {Spencer Jones}, {Fernley},
  {James}, {Fairall}, {Monk}, \& {van Wyk}}]{87aspectra4}
{Whitelock}, P.~A., {et~al.} 1988, \mnras, 234, 5P

\bibitem[{{Whitelock} {et~al.}(1989){Whitelock}, {Catchpole}, {Menzies},
  {Feast}, {Woosley}, {Allen}, {van Wyk}, {Marang}, {Laney}, {Winkler},
  {Sekiguchi}, {Balona}, {Carter}, {Spencer Jones}, {Laing}, {Evans},
  {Fairall}, {Buckley}, {Glass}, {Penston}, {da Costa}, {Bell}, {Hellier},
  {Shara}, \& {Moffat}}]{87aspectra6}
---. 1989, \mnras, 240, 7P

\bibitem[{{Xu} {et~al.}(1995){Xu}, {Crotts}, \& {Kunkel}}]{xu95}
{Xu}, J., {Crotts}, A.~P.~S., \& {Kunkel}, W.~E. 1995, \apj, 451, 806

\end{thebibliography}
\appendix
\section{Effect of Dust Substructure in LE Profiles}
\label{app:substructure}
\begin{figure}[h]
\epsscale{1.1}
\plotone{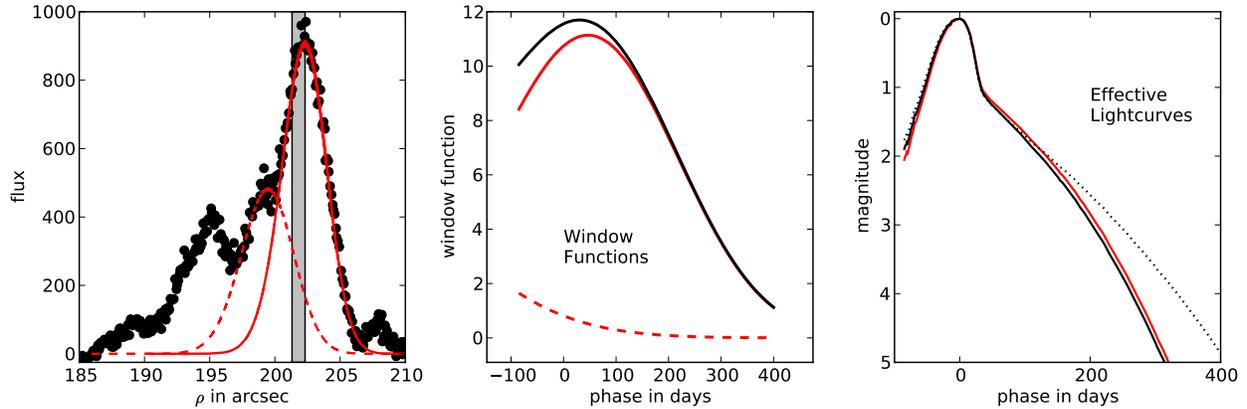}
\caption[]{LEFT: LE113 observed LE profile (black points) with the best fit to the primary dust sheet (solid red line), as well as the best fit to the secondary dust sheet (dashed red line). The location of the slit is shown with the grey region. MIDDLE: Window functions for the primary (solid red line) and secondary (dashed red line) dust sheets, as well as the total window function (solid black line). RIGHT: Effective lightcurves for the primary dust sheet (solid red line), the total effective lightcurve from both dust sheets (solid black line), and the original lightcurve of SN~1987A (dotted line). The effect of the secondary dust sheet is minimal in this case.
\label{fig:f2_substructure}}
\end{figure}

\begin{figure}[h]
\epsscale{1.1}
\plotone{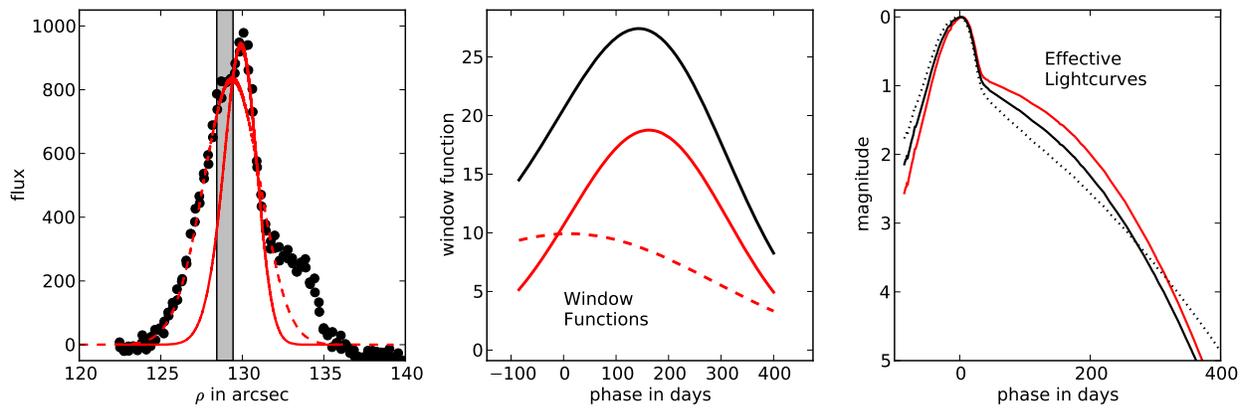}
\caption[]{LEFT: LE186 observed LE profile (black points) with the best fit to the primary dust sheet (solid red line), as well as the best fit to the secondary dust sheet (dashed red line). The location of the slit is shown with the grey region. MIDDLE: Window functions for the primary (solid red line) and secondary (dashed red line) dust sheets, as well as the total window function (solid black line). RIGHT: Effective lightcurves for the primary dust sheet (solid red line), the total effective lightcurve from both dust sheets (solid black line), and the original lightcurve of SN~1987A (dotted line). Without the secondary peak, the effective lightcurve is significantly biased towards later times due to the location of the slit. The secondary peak contributes flux from early epochs, reducing the late-epoch biased produced by the slit.
\label{fig:f3s32_substructure}}
\end{figure}

\begin{figure}[h]
\epsscale{1.1}
\plotone{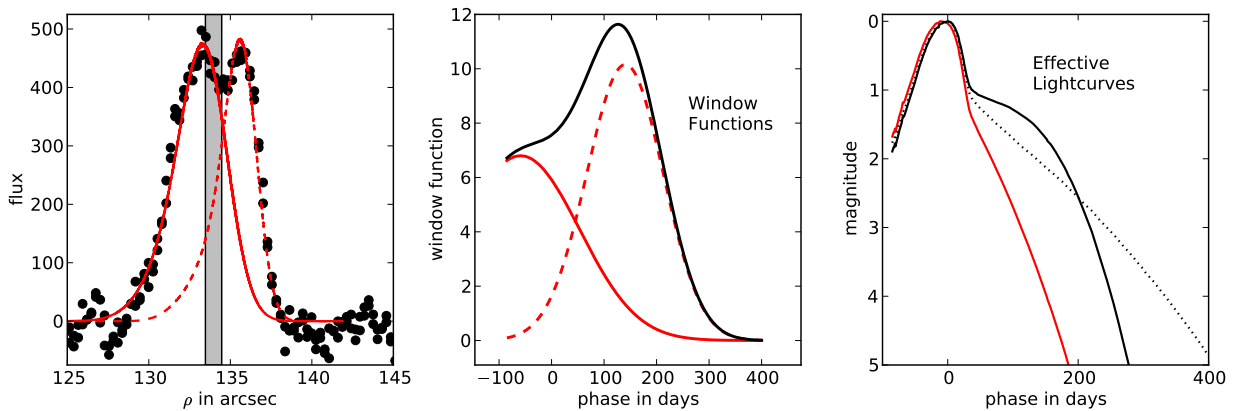}
\caption[]{LEFT: LE180 observed LE profile (black points) with the best fit to
the primary dust sheet (solid red line), as well as the best fit to the
secondary dust sheet (dashed red line). The location of the slit is shown with
the grey region. MIDDLE: Window functions for the primary (solid red line) and
secondary (dashed red line) dust sheets, as well as the total window function
(solid black line). RIGHT: Effective lightcurves for the primary dust sheet
(solid red line), the total effective lightcurve from both dust sheets (solid
black line), and the original lightcurve of SN~1987A (dotted line). Here the two
dust sheets create LE substructure with nearly equal amounts of flux. The window
function in this case is significantly shifted to later times due to the
presence of the secondary peak at larger values of $\rho$.
\label{fig:f3s33_substructure}}
\end{figure}

\begin{figure}[h]
\epsscale{1.1}
\plotone{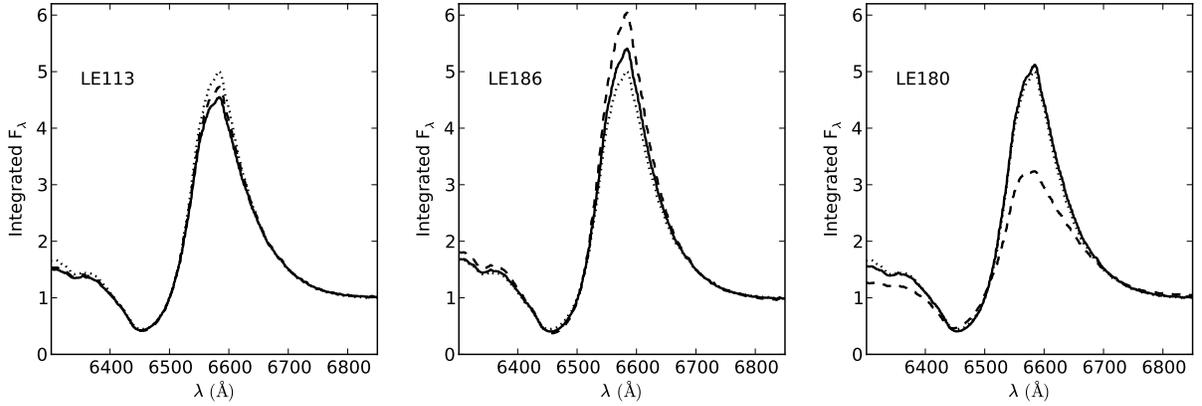}
\caption[]{Resulting \ha profiles obtained by integrating the historical spectra
of SN~1987A with the effective lightcurves in
Figures~\ref{fig:f2_substructure}-\ref{fig:f3s33_substructure} for LE113 (left
panel), LE186 (middle panel), and LE180 (right panel). For each panel, the solid
and dashed lines are from integrations of the total and primary-only effective
lightcurves. The dotted line is \ha profile obtained by integrating the actual
lightcurve of SN~1987A with no window function truncating the integration.
\label{fig:specs_substructure}}
\end{figure}

For the observed LE profiles presented in this work, 11/14 show no strong
evidence for dust substructure within the profile. These LE profiles can be
modeled with one dust sheet effectively. However, LE113, LE180, and LE186 all
show significant substructure in the their profiles
(Figures~\ref{fig:dust_example_f2}~and~\ref{fig:dust_example_f3s32}) and we
seek to quantify the effect this can have on the observed LE spectrum. Figures~\ref{fig:f2_substructure}-\ref{fig:f3s33_substructure} show the observed LE profiles and the effect of dust substructure on the window functions and effective lightcurves. Figure~\ref{fig:specs_substructure} shows the differences in \ha strengths obtained with and without including the effect of the secondary dust structure. 

As shown in Figure~\ref{fig:specs_substructure}, the effect of the secondary
substructure on the resulting \ha profile is minimal for LE113 ($<4\%$
difference in strength of emission peak), primarily because the flux
contribution from the wing of the LE profile is so small. For LE186, including
the secondary LE profile decreases the \ha emission peak by $11\%$. Note,
however, that because of the significant offset of the slit from the primary
peak, the \ha emission for LE186 is in excess of the full lightcurve-weighted
integration of SN~1987A. For the case of LE180, the secondary peak significantly
alters the window function and subsequent \ha line, increasing the emission
strength by $\sim40\%$.
\section{The Excess of \ha Emission in LE186}
\label{sec:app:excess}
The observed LE spectra and isotropic models for LE180 and LE186 in
Figure~\ref{fig:3} show two different results. LE186 shows a clear, high
signal-to-noise excess in \ha emission over the isotropic dust-modeled spectrum.
LE180, probing essentially the same viewing angle onto the photosphere, does not
show such an excess in the \ha profile. In addition to the much lower
signal-to-noise of LE180, we can use arguments based on the above dust
substructure analysis to show that the excess emission observed in LE186 is
physical.

As shown in the middle panel of Figure~\ref{fig:specs_substructure}, the effect of the secondary LE peak in the LE186 profile is to \emph{reduce} the emission peak of \ha. The maximum amount of \ha emission for LE186 therefore occurs when no secondary substructure is included. Since this gives an excess of only $\sim10\%$, even this scenario is not able to account for the $\sim30\%$ excess in \ha in the observed LE186 spectrum.

For the case of LE180, the effect of the substructure is to \emph{increase} the \ha emission peak substantially ($\sim40\%$ using the profile fits in Figure~\ref{fig:f3s33_substructure}). Unlike the case of LE186, where the LE profile fits are well constrained within the data, the fits to the LE180 profile in Figure~\ref{fig:f3s33_substructure} are an example of a situation where our profile fitting algorithms cannot extract meaningful results due to the intrinsic complexity and low signal-to-noise ratio. It is most likely the case the secondary substructure in the LE180 profile is actually much wider than shown in Figure~\ref{fig:f3s33_substructure}, which would result in more late-epoch contribution from the secondary peak and more emission in \ha.
\end{document}